\documentclass[10pt,journal,compsoc]{IEEEtran}
\ifCLASSINFOpdf

\else

\fi
\usepackage{booktabs}
\usepackage{array}
\newcommand{\PreserveBackslash}[1]{\let\temp=\\#1\let\\=\temp}
\newcolumntype{C}[1]{>{\PreserveBackslash\centering}p{#1}}
\newcolumntype{R}[1]{>{\PreserveBackslash\raggedleft}p{#1}}
\newcolumntype{L}[1]{>{\PreserveBackslash\raggedright}p{#1}}
\usepackage{enumerate}
\usepackage{amsmath}
\usepackage{mathtools}
\usepackage{graphicx}
\DeclareGraphicsExtensions{.eps,.ps,.jpg,.bmp}
\usepackage{lineno}
\usepackage{subfigure}
\usepackage{amssymb}
\usepackage{multirow}
\usepackage{algorithm}
\usepackage[noend]{algorithmic}
\usepackage{ifpdf}
\usepackage{verbatim}
\usepackage[numbers,sort&compress]{natbib}

\begin{document}	
	\title{A Unified Collaborative Representation Learning for Neural-Network based Recommender Systems}
	\author{Yuanbo~Xu,
		En~Wang$^\dag$,
		Yongjian~Yang,
		Yi~Chang,
		
		\IEEEcompsocitemizethanks{
			\IEEEcompsocthanksitem Y. Xu, $^\dag$E. Wang (corresponding author) and Y. Yang are with the Department of Computer Science and Technology, Jilin University, Changchun, 130012, China and Key Laboratory of Symbolic Computation and Knowledge Engineering for the Ministry of Education, Jilin University, Changchun, 130012, China. E-mail: {yuanbox, wangen, yyj}@jlu.edu.cn. 
			\IEEEcompsocthanksitem Y. Chang is with the School of Artificial Intelligence, Jilin University, Changchun, Jilin 130012, China. E-mail: yichang@jlu.edu.cn.

			
	}}
	
	%
	%

	\markboth{Journal of \LaTeX\ Class Files,~Vol.~14, No.~8, August~2015}%
	{Shell \MakeLowercase{\textit{et al.}}: Bare Demo of IEEEtran.cls for IEEE Journals}
	%



	\IEEEtitleabstractindextext{%
		\begin{abstract}
			With the boosting of neural networks, recommendation methods become significantly improved by their powerful ability of prediction and inference. Existing neural-network based recommender systems (NN-RSs) usually first employ matrix embedding (ME) as a pre-process to learn users' and items' representations (latent vectors), then input these representations to a specific modified neural network framework to make accurate Top-k recommendations. Obviously, the performance of ME has a significant effect on RS models. However, most NN-RSs focus on accuracy by building representations from the direct user-item interactions (e.g., user-item rating matrix), while ignoring the underlying relatedness between users and items (e.g., users who rate the same ratings for the same items should be embedded into similar representations), which is an ideological disadvantage. On the other hand, ME models directly employ inner products as a default loss function metric that cannot project users and items into a proper latent space, which is a methodological disadvantage. In this paper, we propose a supervised collaborative representation learning model - Magnetic Metric Learning (MML) - to map users and items into a unified latent vector space, enhancing the representation learning for NN-RSs. Firstly, MML utilizes dual triplets to model not only the observed relationships between users and items, but also the underlying relationships between users as well as items to overcome the ideological disadvantage. Specifically, a modified metric-based dual loss function is proposed in MML to gather similar entities and disperse the dissimilar ones. With MML, we can easily compare all the relationships (user to user,	item to item, user to item) according to the weighted metric, which overcomes the methodological disadvantage. We conduct extensive experiments on four real-world datasets with large item space. The results demonstrate that MML can learn a proper unified latent space for representations from the user-item matrix with high accuracy and effectiveness, and lead to a performance gain over the state-of-the-art RS models by an average of 17\%.
		\end{abstract}
		
		\begin{IEEEkeywords}
			Latent vectors, Collaborative  Representation Learning, Metric Learning, Recommender Systems.
	\end{IEEEkeywords}}

	\maketitle

	%
	\IEEEpeerreviewmaketitle

	\section{Introduction}
	
	In recent years, popular online commercial websites such as Netflix, Amazon, Yelp, and Taobao provide a wide spectrum of recommendation services to help the customers filter their preferences out of enormous product space \cite{101DBLP:conf/www/PatroBGGC20}. However, the performance of traditional recommendation models, such as collaborative filtering (CF) \cite{102DBLP:conf/recsys/Fan0YWTL19}, matrix factorization (MF) \cite{103DBLP:conf/ijcai/BugliarelloJR19} is highly restricted by the large scale of product space. With the development of neural networks and computation theory, the technology of recommender systems has been taken to the next stage \cite{11ricci2015recommender}. To tackle large scale products for recommendations, most neural-network-based recommender systems first extract latent vectors of users and items from a user-item matrix. This extraction procedure is called matrix embedding (ME) \cite{12zhang2017deep}, which is a critical factor in getting accurate recommendations, especially for learning meaningful, measurable latent vectors. With these latent vectors, some traditional recommendation models are enhanced for real-world applications, such as CF to NCF \cite{5he2017neural}, MF to NeuMF \cite{106DBLP:journals/tnn/FanW17}. Some novel NN-based recommendation models are also proposed, such as GERL \cite{104DBLP:conf/www/GeWWQH20}, NeuO \cite{105DBLP:journals/nn/XuYHWZYX19} and HERec \cite{118DBLP:journals/tkde/ShiHZY19}.
	
	\begin{figure}[tbp]
		\centering
		\includegraphics[width=0.8\columnwidth]{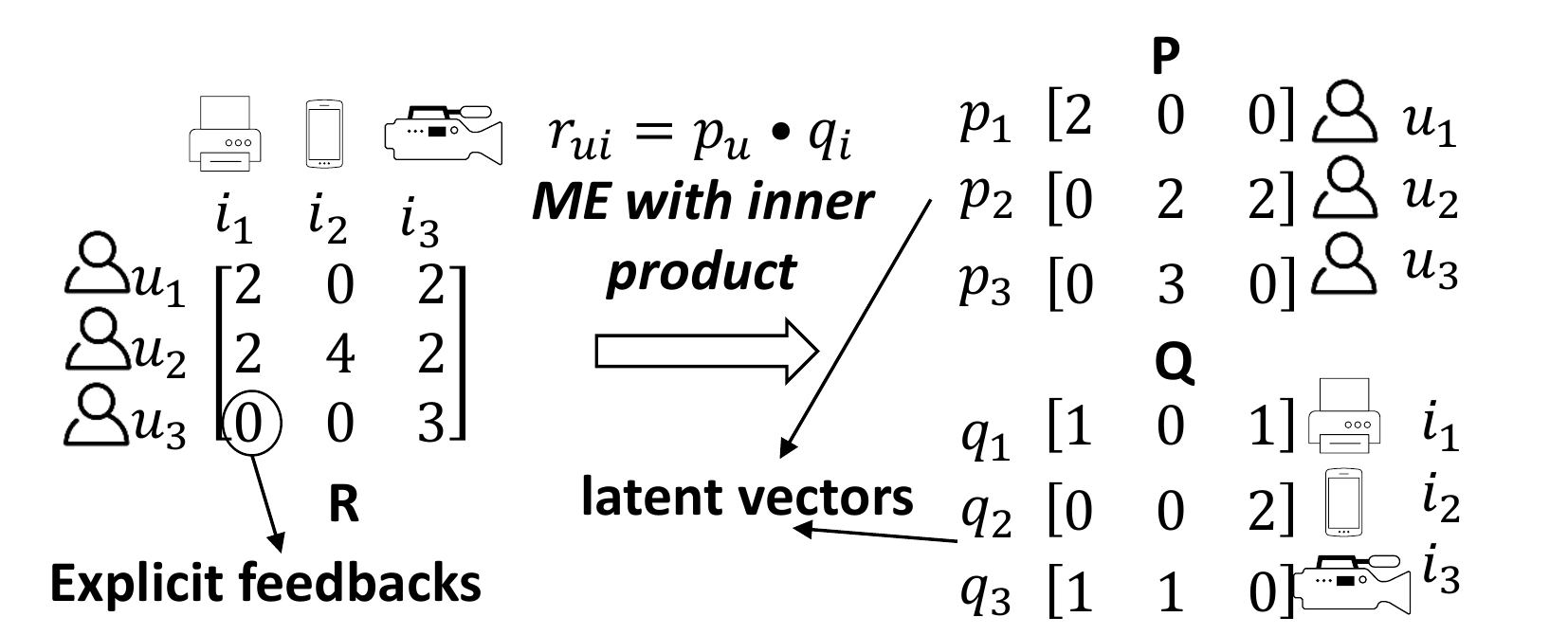}
		\caption{An example to illustrate the disadvantages of traditional ME models.}
		\label{Fig:MME1}
		\vspace{-10pt}
	\end{figure}
	
	However, most researches assume that these latent vectors learned by existing ME models are insufficient and biased \cite{11ricci2015recommender}, without taking the interpretability into consideration \cite{11ricci2015recommender,14he2016interactive}. In other words, traditional ME only utilizes the relationships between users and items, while ignoring that the collaborative relationships between users and users, items and items, which is an \textbf{\textit{ideological disadvantage}}. Moreover, most existing works directly utilize inner products to measure the relationships between users and items. This simple metric may cause chaos when computing similarities, which is a \textbf{\textit{methodological disadvantage}}. 
	
	To make the above two disadvantages clear, we give a recommendation scenario in Fig.\ref{Fig:MME1}, where we employ basic matrix factorization as ME model and user-based collaborative filtering (UBCF \cite{2hu2008collaborative}) as recommendation model. In this example, $U (u_1,u_2,u_3)$ and $I (i_1,i_2,i_3)$ represent users and items, respectively. $R$ is a user-item rating matrix with ratings $r_{ij}$. $P$ and $Q$ are built with 3-dimensional embedding results ($p_i$ for user $i$, $q_j$ for user $j$), extracted from matrix $R$ with existing ME models, such as SVD or other matrix factorization methods. To pick a proper item recommended to $u_2$, we employ a popular recommendation model (user-based CF, UBCF) with users' latent vectors. User-based CF calculates the similarities among $u_1,u_2,u_3$ to pick the Top-1 user neighbor for $u_2$. Then it recommends items that this Top-1 user has consumed to $u_2$. 
	
	As a result in Fig.\ref{Fig:MME1}, in ideology (note that UBCF model uses inner products to calculate latent vectors while the similarity between latent vectors is measured by Euclidean distance \cite{14he2016interactive}). Intuitively, UBCF should pick $u_3$ as $u_2$'s Top-1 neighbor because $(p_1, p_2)_E=\sqrt {{\rm{12}}}  > (p_3, p_2)_E=\sqrt {{\rm{5}}}$. But in fact, when considering the underlying relationship (between $u_1$ and $u_2$) hidden in matrix $R$, it is obvious that $u_1$ should be a better choice ($u_1$ and $u_2$ share the same preference of $i_1$ and $i_3$ according to their explicit feedbacks $r_{11}$, $r_{13}$, $r_{21}$ and $r_{23}$). Hence, choosing $u_3$ as $u_2$'s neighbor is an inaccurate decision caused by the \textit{\textbf{ideological}} \textbf{\textit{disadvantage}}, which is partly mentioned in \cite{1hsieh2017collaborative,121DBLP:conf/aaai/LiZZQZHH20}.
	
	In methodology, directly choosing inner products as the metric may cause a dilemma, especially for CF models. In general, CF models employ Euclidean distance between latent vectors as the similarity to find the nearest neighbor \cite{11ricci2015recommender}, where the latent vectors are learned by using inner products in traditional ME models \cite{107DBLP:conf/aaai/AcharyaGMD19}. To ensure the metric-satisfying non-negativity in latent space, the latent vector calculation should obey the triangle inequality (the sum length of any two sides must be greater than or equal to the remaining side, and the reason why embeddings should obey this is detailed introduced in \cite{5he2017neural,14he2016interactive}). However, the relationships measured by inner products may violate the triangle inequality. For example, as shown in Fig.\ref{Fig:MME1}, if ME models use inner products $\bullet$ to learn latent vectors for $i_1,u_2$ and $u_3$ as $q_1,p_2$ and $p_3$, then, $p_2 \bullet q_1+p_3 \bullet q_1 < p_2 \bullet p_3$, which violates the triangle inequality. If we conduct calculations in a latent space with a metric that violates the triangle inequality, it may lead to uncertainty and inaccuracy of computing, and finally, result in a biased recommendation. Therefore, only applying inner products in ME models is not a suitable choice when learning latent vectors for recommendations. This \textit{\textbf{methodological}} \textbf{\textit{disadvantage}} damages the performance of recommendation models tremendously. For recommender systems, it's still a challenge to learn a proper latent space, where all kinds of relationships (users/items/user-item) can be measured by a unified style of the metric.

	\begin{figure}[tbp]
		\centering
		\includegraphics[width=0.8\columnwidth]{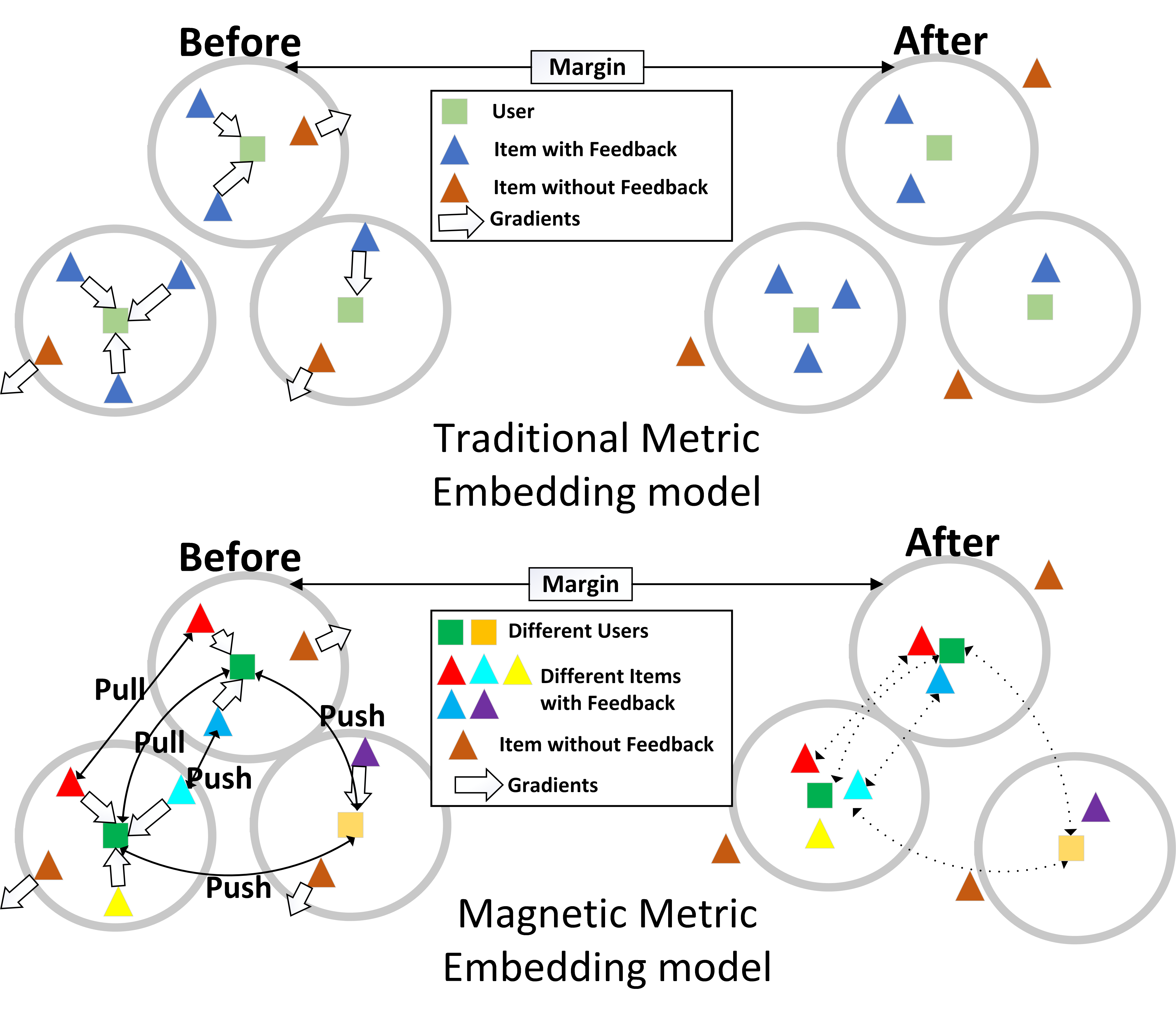}
		\caption{Comparison between traditional metric learning and our proposed model MML. Traditional ML (upper part) only focuses on user-item relationships (rectangle to triangle), while MML (bottom part) also takes the underlying user-user (rectangle to rectangle), item-item (triangle to triangle) relationships into consideration.}
		\label{Fig:MME2}
		\vspace{-10pt}
	\end{figure}
	
	To relieve the limitation of inner products, metric learning has been proved to be useful in the multimedia area \cite{1hsieh2017collaborative,20Wang2018Robust,108DBLP:journals/kais/NikolakopoulosK19}. The core of metric learning is to learn a proper metric for the measurement between latent vectors. However, metric learning is only designed to measure user-item relationships in recommender systems \cite{121DBLP:conf/aaai/LiZZQZHH20, 1hsieh2017collaborative}, which cannot simultaneously tackle the \textit{\textbf{ideological}} and \textit{\textbf{methodological}} disadvantage (as shown in Fig.\ref{Fig:MME2}). To this end, we propose a supervised collaborative representation learning model for matrix embedding: Magnetic Metric Learning (MML), which utilizes the dual triplets to represent the different types of relationships (user-user, item-item, user-item) with a uniform latent space in a uniform framework. MML can learn not only the explicit relationships but also the latent relationships, which overcomes the ideological disadvantages. Meanwhile, the relationships between users and items are directly measured by weighted metric distance, which overcomes the methodological disadvantage. 
	
	The contributions of this paper are summarized as follows:
	\begin{itemize}
		\item We first argue that existing matrix embedding methods for neural-network-based recommendation models are not sufficient and unbiased. Then we explore the ideological and methodological disadvantages of traditional ME models and propose a representation learning model for matrix embedding: Magnetic Metric Learning, to overcome the above disadvantages.
		\item For the ideological disadvantage, we utilize dual triplets to model explicit and latent collaborative relationships among users and items in a uniform latent space. For the methodological disadvantage, a modified metric-based dual loss function is proposed to learn weighted metric and latent vectors at the same time. 
		\item The experimental results on four real-world datasets demonstrate that MML can learn a proper unified latent space from the user-item matrix, and improve the accuracy of the state-of-the-art models. 
	\end{itemize}
	
	The paper is organized as follows. We provide preliminaries in Section 2. Then we elaborate on the proposed method MML, including theory, regularization, and training process in Section 3. We report the experimental results in Section 4. Lastly, we review related work in Section 5 and conclude this paper in Section 6.
	
	\section{Preliminaries}
	\subsection{Basic definitions} 
	In recommender systems, $ U $ denotes a set of $ m $ users $ U=\{u_1,u_2...u_m\} $, and $ I $ denotes a set of $ n $ items $ I=\{i_1,i_2...i_n\} $. A user-item rating matrix, whose entries are $r_{ui}$, is built as $ R $. For items with ratings, we set $r_{ui}$ as the rating, while for items without ratings, $r_{ui}=0$. If $r_{ui}=0$, we treat $(u,i)$ as a negative pair, otherwise a positive pair. 
	
	\textbf{Definition 1: Matrix Embedding: } Given a matrix $R \in \mathbb{R}^{m\times n}$ , the matrix embedding (ME) model is to depose the matrix into two low-dimension $k$ spaces, which are also called latent vector spaces $P \in \mathbb{R}^{m\times k}$, $Q \in \mathbb{R}^{n\times k}$. Especially, in recommender systems, $R$ is the user-item rating matrix. $p_i \in P$ is the latent vector for user $i$, while $q_j \in Q$ is for item $j$. Note that in real-world scenarios, the scale of users and items is huge, which means that $m, n \gg k$. Existing ME models usually utilize some matrix factorization methods, such as pureSVD \cite{108DBLP:journals/kais/NikolakopoulosK19} and NMF \cite{108DBLP:journals/kais/NikolakopoulosK19} to learn latent vectors. However, these models directly employ inner products in their loss function, which may lead to inaccurate and biased embedding results. Our proposed model aims to solve this problem, which is demonstrated in detail in Section 3.
	
	\textbf{Definition 2: Metric Learning: } Given two different latent vectors $p,q \in \mathbb{R}^{1 \times k}$, the metric learning (ML) model is to learn a proper weighted metric matrix $W \in \mathbb{R}^{k\times k}$ to measure the relationships between $p$ and $q$ \cite{111DBLP:conf/aaai/LiZZQZHH20}. The different weights in $W$ stand for the importance of each element in latent vectors. Specifically, in recommender systems, the distance between $p_i$ and $q_j$ can be treated as the measurement between user $i$ and item $j$, as well as the user $i$'s preference for item $j$. Existing metric learning models usually focus on the explicit feedbacks and models' optimizations in recommender systems, such as CML \cite{1hsieh2017collaborative} and IML \cite{6ijcai2018-389}. However, these models usually ignore the underlying relationships hidden in the user-item matrix $R$, which is a restriction to the ML models' performance.
	
\begin{figure}[tbp]
		\centering
		\includegraphics[width=0.9\columnwidth]{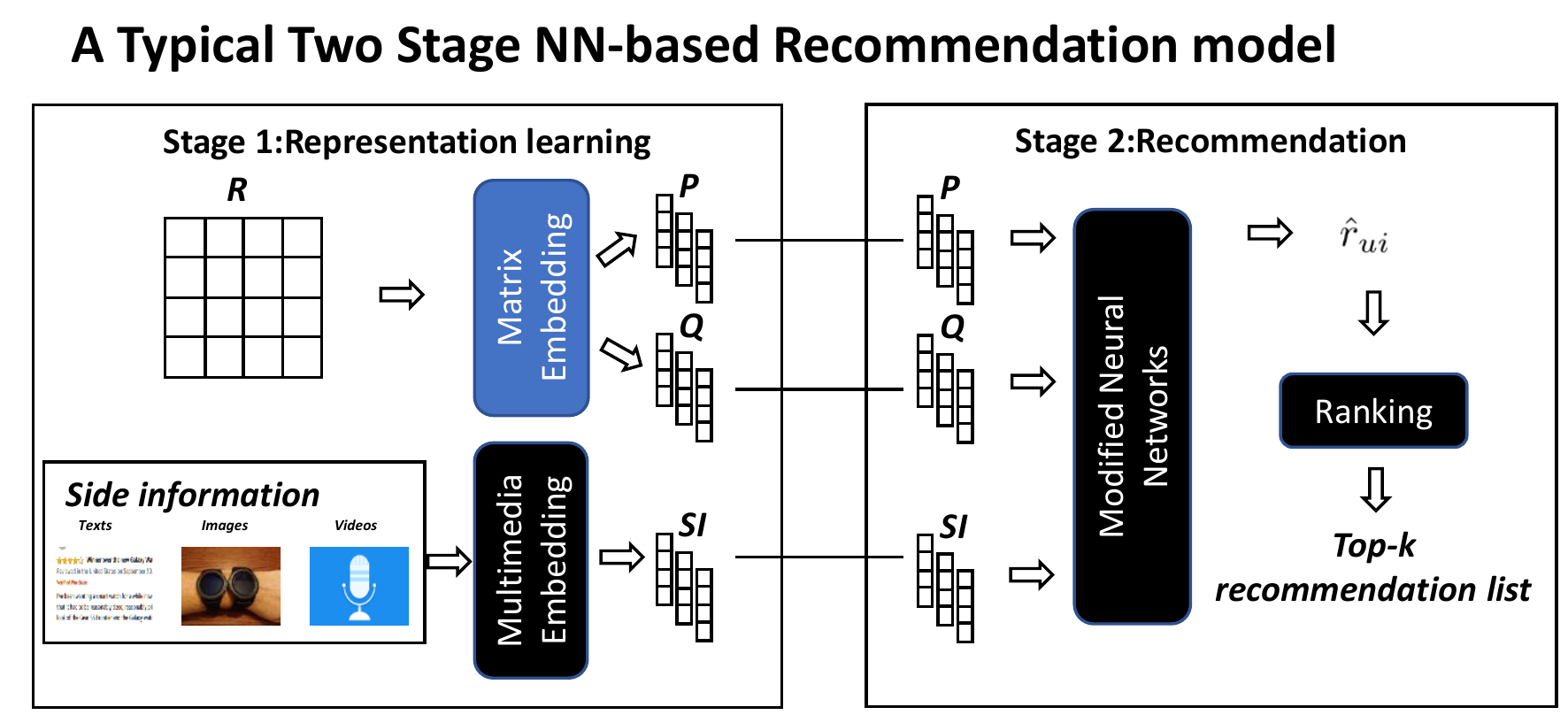}
		\caption{A typical two-stage neural network based recommendation model. Note that in this paper we focus on the matrix embedding part of this framework.}
		\label{Fig:MME11}
	\end{figure}

	\textbf{Definition 3: Neural-Network-based Recommendation models: } A typical neural-network-based recommendation model is a two-stage framework: the basic input is the user-item rating matrix $R$, and some other side information, including text, videos, and images. The first stage is named representation learning. In this stage, the inputs are mapped into latent vectors, including user latent vectors $P$, item latent vectors $Q$, and side information latent vectors $SI$, which extracts the latent features hidden in the multi-modal information. In the second stage for the recommendation, the latent vectors are feed into a modified neural network, which outputs the predicted ratings $\hat r_{ui}$. Finally, according to the ranking of $\hat r_{ui}$, the model gives a Top-k recommendation list. A general framework is shown in Fig.\ref{Fig:MME11}. Some popular recommendation models are based on this framework with different embedding models and neural networks, including Neural CF \cite{5he2017neural}, NeuO \cite{105DBLP:journals/nn/XuYHWZYX19}. However, most models use inner products as default, where we argue it is not always stable.
	
	Note that some existing models integrate and implement joint learning framework. However, we argue there are some disadvantages: 1) overfitting. If we co-train the two stages, we have only one loss function on the recommedation stage, which may lead to the potential overfitting problem in representation learning stage \cite{117DBLP:conf/www/MaZYYCTHC20}. 2) flexibility. The users' and items' latent representations learned by first stage could be combined with different recommendation model, or other models (such as user profiling, slanderous user detection), which is flexible for different application scenarios.
	\subsection{Matrix embedding with inner products}
	Given a user-item matrix $R$, matrix embedding models with inner products usually minimize this loss function $L_{\text{IP}}$ to learn latent vectors $P$ and $Q$:
	\begin{equation}
	\begin{aligned}
	{L_{\text{IP}}} = \sum\limits_{u \in U,i \in I}^{P,Q} {({r_{ui}}}  - {p_u} \bullet {q_i}{)^2} + pen(P,Q),
	\end{aligned}
	\label{eq1} 
	\end{equation}
	where $pen(P,Q)$ is a penalty term to avoid overfitting. Then we can use these latent vectors to make recommendations: 1) for user-based collaborative filtering \cite{26koren2015advances}, we need to find the nearest k-neighbor for target user $t$ with the following function: 
	\begin{equation}
	\begin{aligned}
	{N_k}(\mathop {\min }\limits_{u \in U} {\left| {{p_u},{p_t}} \right|_{Euc}}).
	\end{aligned}
	\label{eq2} 
	\end{equation}
	
	Then some common items in these neighbors can be recommeded to the target user. 2) for neural network based models, we input the latent vectors and make recommedations as shown in Fig.\ref{Fig:MME11}.
	\subsection{Matrix embedding with metric learning}
	Given a user-item matrix $R$, matrix embedding models with metric learning usually minimize this loss function $L_{\text{ML}}$:
	\begin{equation}
	\begin{aligned}
	{L_{\text{ML}}}\substack{=}\sum\limits_{u \in U,{i,j} \in I}^{P,Q} {\mathop {({L_{\text{pull}}}}\limits_{{r_{ui}} \ne 0} ({p_u},{q_i}) \substack{-} \mathop {{L_{\text{push}}}}\limits_{{r_{uj}} = 0} ({p_u},{q_j})) \substack{+} \text{pen}(P,Q)}.
	\end{aligned}
	\label{eq3} 
	\end{equation}
	
	Note that there are two loss functions in traditional metric learning: $L_\text{pull}$ and $L_\text{push}$. The core idea for metric learning is to gather the user-item pair with explicit feedbacks and disperse the pair without them. So $L_{\text{pull}}$ is employed to calculate the weighted distance between user $u$ and item $i$, where $r_{ui}\ne 0$. By minimizing $L_{\text{pull}}$, $L_{\text{ML}}$ tries to pull the user and item together. Meanwhile, by minimizing $-L_{\text{push}}$, $L_{\text{ML}}$ tries to push away the user and item where $r_{ui} = 0$. Specifically, in the training process of metric learning, the model can learn not only the latent vectors $P$,$Q$, but also the weighted metric matrix $W$. The important notations are shown in Table \ref{TKtable1}.
	
	\begin{table}[htbp]
		\centering
		\caption{Notation List.}
		\resizebox{0.4\textwidth}{!}{
			\begin{tabular}{l|l}
				\hline\hline
				Notation & Description \\
				\hline
				$ U $ & user set with $u$ in recommender systems\\\hline
				$ I $ & item set with $i$ in recommender systems\\\hline
				$ R $ & rating matrix $R \in \mathbb{R}^{m \times n}$ with rating $r_{ui}$\\\hline
				$ \mathbb{R}, \mathbb{E}$ & notations for latent spaces\\\hline
				$m, n$ & number of users/items\\\hline
				$ r_{ui} $ & $ u $'s rating on item $ i $ \\\hline
				$ P $ & $k$-dimension user latent vector set $P \in \mathbb{R}^{m \times k}$\\\hline
				$ Q $ & $k$-dimension item latent vector set $Q \in \mathbb{R}^{n \times k}$\\\hline
				$ p_u, q_i $ & $k$-dimension latent vectors for $u$ and $i$\\\hline
				$ e $ & uniformed latent vector (a user or an item)\\\hline
				$ W $&  metric matrix $W \in \mathbb{R}^{k \times k}$ with $w$\\\hline
				$ W^U, W^I, W^{UI} $& $W$ for users, items and user-item\\\hline
				$\left|{e_a,e_b} \right|_\text{Euc}$ & Euclidean distance between ${e_a,e_b}$\\\hline
				$\left|{e_a,e_b} \right|_\text{W}$ & metric distance between ${e_a,e_b}$ with $W$\\\hline
				${L^\text{EX}_\text{MML}}$ & explicit relationship loss function \\\hline
				$ S^{U},S^{I}$ & user/item similar-pair buffer sets \\\hline
				$ mr $ & learning margin for metric learning \\\hline
				${L^\text{LA}_\text{MML}}$ & latent relationship loss function \\\hline
				$\alpha, \lambda, \theta, \omega $ & hyper parameters \\\hline
				\hline
		\end{tabular}}
		\label{TKtable1}%
	\end{table}%
	
	\section{Magnetic Metric Learning (MML) model}
	Magnetic Metric Learning model (MML) employs a unified style of metric learned through embedding and recommendation, and learns a unified latent space for users and items, which overcomes the methodological disadvantage. Meanwhile, MML considers both explicit and latent relationships and makes a direct embedding to overcome the ideological disadvantage (shown in Fig.\ref{Fig:MME2}).
	
	Specifically, MML treats users and items as the same entities in a unified latent space, where all the relationships between users and items can be represented by their distance (in MML, it is measured by learned metric $W$). Moreover, MML can learn users’ and items’ latent vectors in a uniform framework with a uniform metric across all the procedures (embedding and recommendation) and overcome the limitation of inner products. MML does not need to distinguish latent user space and latent item space. All the users and items are embedded into the same dimension latent space. In this way, we could more easily optimize MML’s loss function and calculate its gradient compared with other NN-based embedding models, such as autoencoder. 
	\subsection{Learning Metric: Foundation of MML}
	
	We define a $k$-dimensional uniformed latent space $\mathbb{E}$, where $e_i \in \mathbb{E}$ stands for an extracted latent vector for a user or an item, $i$ stands for an entity which can be either a user or an item. First, we define the function $F$ for calculating the relationships between entities $a,b$ as the following Euclidean function:
	
	\begin{equation}
	\begin{aligned}
	{F^E}(a,b) = \left\| {{e_a} - {e_b}} \right\|_{\text{Euc}}^2.
	\end{aligned}
	\label{eq4} 
	\end{equation}
	
	While in MML, we use a learned metric $W \in \mathbb{R}^{k \times k}$ as a substitute for Euclidean, as shown in Eq.(\ref{eq14}):
	
	\begin{equation}
	\begin{aligned}
	{F^*}(a,b) = \left\| {{e_a} - {e_b}} \right\|_{W^*}^2.
	\end{aligned}
	\label{eq14} 
	\end{equation}
	
	Note that we consider learning different weighted matrix $W^*$ for measuring user-user  ($W^{U}$), item-item ($W^{I}$) and user-item ($W^{UI}$) relationships, which is an improvement over other metric learning models, such as CML \cite{1hsieh2017collaborative}, IML \cite{6ijcai2018-389} and CRML \cite{112DBLP:journals/nn/WuZNC20}. With these learned metrics, all the relationships can be measure as follows:
	\begin{equation}
	\begin{aligned}
	\left\| {{e_a} - {e_b}} \right\|_{{W^*}}^2{\rm{ = }}\sqrt {{{({e_a} - {e_b})}^T}{W^*}({e_a} - {e_b})}. 
	\end{aligned}
	\label{eq15} 
	\end{equation}
	
	To ensure that the $W^*$ we learned is a metric-satisfying non-negative metric and obeys the triangle inequality, we need to require $W^*$ to be positive semi-definite. Note that setting $W^*$=$I$ gives Euclidean distance. And if we set $W^*$ to be diagonal, it corresponds to learning a metric in which different axes are given different weights upon Euclidean distance. Generally, $W^*$ parameterizes a family of Mahalanobis distance over $\mathbb{R}^{k \times k}$ \cite{9xing2003distance,112DBLP:journals/nn/WuZNC20}. With different restrictions to $W^*$, we can tune our proposed model MML for different application scenarios.
	\subsection{Explicit relationships formulation}
	
	MML is designed to gather similar entities and disperse the dissimilar ones with learned metrics. In recommender systems, we treat the feedback $r_{ui} \in R$ as the indicator of explicit relationships. If $r_{ui}\ne 0$, we define that there is an explicit relationship between user $u$ and item $i$. To consider this for enhancing matrix embedding process, we sample the dual triplets $<a,b,c>$ and $<c,d,a>$, where $a,b \in U$, $c,d \in I$, and $r_{ac} \ne 0, r_{bc}=0, r_{ad}=0$. To ensure the structural consistency, we can learn that $a,c$ should be embedded closer than $b,c$ and $a,d$. Meanwhile, according to $a,b$'s different preferences on $c$, it is obvious that they should not be embedded closely. The same deduction is applied on ($b,c$), ($c,d$) and ($a,d$). In this way, MML maximizes the effect of metric learning with a modified enhanced metric-based dual loss function (EMDL):
	\begin{equation}
	\begin{aligned}
	&L_\text{MML}^1 =\\ 
	&\sum\limits_{a,b \in U;c \in I}{{t_{a,b,c}}|{mr_1}\substack{+}F^{UI}(e_a,e_c)\substack{-}F^U(e_a,e_b)\substack{-}F^{UI}(e_b,e_c){|_{\substack{+}}}},
	\end{aligned}
	\label{eq5} 
	\end{equation}
	\begin{equation}
	\begin{aligned}
	&L_\text{MML}^2 = \\
	&\sum\limits_{a \in U; c,d \in I} {{t_{a,c,d}}|{mr_2}\substack{\textsc{+}}F^{UI}(e_a,e_c)\substack{-}F^{I}(e_c,e_d)\substack{-} F^{UI}(e_a,e_d){|_ + }},
	\end{aligned}
	\label{eq6} 
	\end{equation}
	where notation $|J|_+$ satisfies that: $|J|_+ = max(J; 0)$. $t$ is a
	ranking weight calculated as suggested in \cite{1hsieh2017collaborative}. And $mr_1,mr_2>0$ is the safety margin size. With this dual loss function, ($a,c$) is embedding closer than ($a,d$), ($b,c$) with metric $ W^{UI}$. ($a,b$) and ($c,d$) are embedded far with metric $W^U$ and $W^I$. Finally, we get the EMDL loss function of MML for explicit relationships:
	\begin{equation}
	\begin{aligned}
	{L^\text{EX}_\text{MML}} = \lambda L_\text{MML}^1 + (1 - \lambda )L_\text{MML}^2,
	\end{aligned}
	\label{eq7} 
	\end{equation}
	where $\lambda$ is the balance weight between users and items. By minimizing EMDL, we can not only pull the user-item pair together with explicit relationships ($r_{ac} \ne 0$) and push away the user-item pairs with no feedback ($r_{bc},r_{ad} = 0$), but also push away the user-user pair ($a,b$) and item-item pair ($c,d$), as shown in the lower part in Fig.\ref{Fig:MME2}. 
	\subsection{Latent relationships formulation}
	Different from explicit relationships between users and items which are indicated by $r_{ui} \in R$, latent relationships always occur between users and users, items and items, which can not be directly observed. So many existing matrix embedding models only consider explicit relationships while ignoring the latent ones. However, the latent relationships should be an important factor in matrix embedding because they also reflect the users' preferences and items' features, as the example we have given in Introduction. In order to utilize the latent relationships, we first extract the user pairs and item pairs according to the following rules:
	\begin{itemize} 
		\item Users who rate the same items should be embedded closer in latent vector space, and vice versa.
		\item Items rated by the same users should be embedded closer in latent vector space, and vice versa.
	\end{itemize}
	
	With the rules above, we first build two similar-pair buffer sets: $S^U$ and $S^I$, which contain user pairs and item pairs, respectively. We treat user pair in $S^U$ as the same category, so do the item pair in $S^I$. A user-user or item-item pair $(a,b)$ is assigned to similar-pair buffer sets according to the following restriction:
	
	\begin{equation}
	\begin{aligned}
	\left\{ {\begin{array}{*{20}{c}}
		{({a},{b}) \in {S^{(*)}}, \text{if}~\frac{{|\text{list}(a) \cap \text{list}(b)|}}{{|\text{list}(a) \cup \text{list}(b)|}} > {\theta };} \\ 
		{({a},{b}) \notin {S^{(*)}},       \text{else},} 
		\end{array}} \right.
	\end{aligned}
	\label{eq8} 
	\end{equation}
	where $\text{list}(a)$ means the list of items that user $a$ has rated, or the users who have rated item $a$, and ${S^{(*)}}$ is either ${S^U}$ or ${S^I}$. $\theta$ is a control threshold to decide the partition of same preference that the users or the items share. So the loss function of latent relationships is as follows:
	
	\begin{equation}
	\begin{aligned}
	L_\text{MML}^\text{LA} = \sum\limits_{a,f \in {S^{(*)}}} {\sum\limits_{a,g \notin {S^{(*)}}} {{t_{a,f,g}}|{mr_3} + F(e_a,e_f) - F(e_a,e_g){|_ + }} }.
	\end{aligned}
	\label{eq9} 
	\end{equation}
	
	In Eq.(\ref{eq9}), $f$ is the similar entity of $a$, while $g$ is not. $F$ could be either $F^{U}$ or $F^{I}$ in one formulation. $mr_3$ is the safety margin size. With this formulation, the user pair or item pair ($a,f$) in $S$ are embedded closer than ($a,g$) not in $S$. The matrix embedding is more enhanced by considering the latent relationships for both users and items.
	\subsection{Magnetic Metric Learning Formulation}
	Finally, we combine explicit relationship loss $L^\text{EX}_\text{MML}$ and latent relationship loss $L^\text{LA}_\text{MML}$ linearly with a combination weight $\alpha$:
	
	\begin{equation}
	\begin{aligned}
	{L_\text{MML}} = \alpha L_\text{MML}^\text{EX} + (1 - \alpha )L_\text{MML}^\text{LA}.
	\end{aligned}
	\label{eq10} 
	\end{equation}
	
	Note that in $L_\text{MML}^\text{EX}$ and $L_\text{MML}^\text{LA}$,  all the +$F$ functions are the realizations of $L_{pull}$ in Eq.(\ref{eq2}), which means pulling the similar entities together in learned metric space. While the -$F$ functions mean $L_{push}$, which pushes the dissimilar entities away.
	\subsection{Regularization and Optimization}
	We add two regularizations to make MML efficient and feasible. 
	
	To avoid overfitting and biased parameters, we bound all the embedding results $e^{(*)}$ (users' and items' latent vectors) in a unit sphere: $||{{e}^{(*)}}|{|^2} < 1$, to ensure the robustness of our model.
	
	Moreover, we utilize a covariance regularization proposed by \cite{25cogswell2015reducing} to restrict the embedding results. First, we calculate a $k \times k$ matrix $E$ for an $O$ size of $k$-dimension vector $e$: 
	
	\begin{equation}
	\begin{aligned}
	{E_{ij}} = \frac{1}{O}\sum\limits_o {(e_i^o - {\eta _i})} (e_j^o - {\eta _j}),
	\end{aligned}
	\label{eq16} 
	\end{equation}
	where $o$ denotes the index in $O$, $i, j$ is an index pair in a range of $k$. ${\eta _i} = \frac{1}{O}\sum\limits_o {e_i^o}$. Then we define penalty loss $L_P$:
	
	\begin{equation}
	\begin{aligned}
	{L_\text{P}} &= \frac{1}{O}({\left\| E \right\|_\text{f}} - \left\| {\text{diag}(E)} \right\|_2^2); \hfill \\
	&\text{Subject to}~~ ||{e^{(*)}}|{|^2} < 1, \hfill \\
	\end{aligned}
	\label{eq11} 
	\end{equation}
	where ${\left\| E \right\|_\text{f}}$ is F-norm of $E$, $ {\text{diag}(E)}$ is a diagonal matrix.
	
	Moreover, to optimize the model, we first define the user-user weighted metric matrix $W^U$ and item-item weighted metric matrix $W^I$ to be symmetric because the relationships among users or items are undirected. With this restriction, we can save up the running time when calculating gradients. 
	
\begin{figure}[htbp]
		\centering
		\includegraphics[width=0.8\columnwidth]{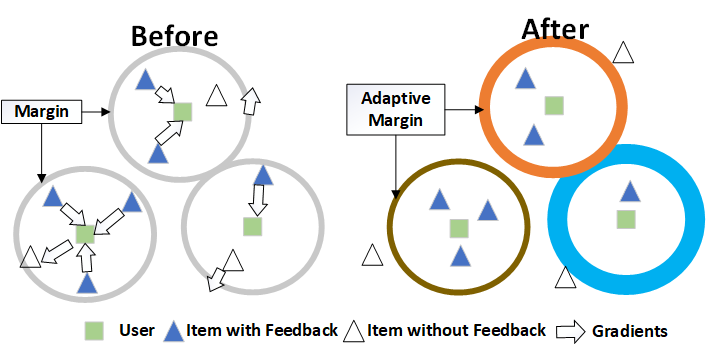}
		\caption{Effect of applying adaptive margins in MML.}
		\label{Fig:MME16}
	\end{figure}

	To add personality into our model, we employ adaptive margins in MML (Fig.\ref{Fig:MME16}). Specially, there are three different margins in our model: $mr_1$, $mr_2$ and $mr_3$. Inspired by \cite{121DBLP:conf/aaai/LiZZQZHH20}, we prefer to use adaptive margin to reduce the variations, which utilizes $mr^u$, $mr^i$ and $mr^l$ to replace the origin margin $mr_1$, $mr_2$ and $mr_3$, respectively for different categories of relationships. Note that the less items the users have rated, the larger margin should be applied to avoid overfitting. Thus, the adaptive margins could be achieved by minimizing the following loss function $L_\text{R}$: 
	
	\begin{equation}
	\begin{aligned}
	{L_\text{R}} &=  - (\frac{1}{m}\sum\limits_u {mr^u}  + \frac{1}{n}\sum\limits_u {mr^i}  + \frac{1}{{m + n}}\sum\limits_u {mr^l} );\hfill \\
	&\text{Subject to}~~ mr^u \in (0,1], mr^i \in (0,1], mr^l \in (0,1], \hfill \\
	\end{aligned}
	\label{eq17} 
	\end{equation}
	where $m, n$ are the size of $U$, $I$.
	
	\subsection{Training process}
	In summary, our complete loss function of MML is shown as follows:
	\begin{equation}
	\begin{aligned}
	&~~~~~~~~\mathop {\text{Minimize} }\limits_{{e^{(*)}}}( {L_\text{MML}} + {\omega _\text{P}}{L_\text{P}}+ {\omega _\text{R}}{L_\text{R}}); \hfill \\
	&~~~~~~~~~\text{Subject to}~~ ||{e^{(*)}}|{|^2} < 1, \hfill \\
	&~~~~~~~~~~~~~~~~~~~~~~~~~~~mr^u \in (0,1], \hfill \\
	&~~~~~~~~~~~~~~~~~~~~~~~~~~~mr^i \in (0,1], \hfill \\
	&~~~~~~~~~~~~~~~~~~~~~~~~~~~mr^l \in (0,1], \hfill \\
	\end{aligned}
	\label{eq12} 
	\end{equation}
	where $\omega_\text{P}$, $\omega_\text{R}$ are the hyperparameters for controlling $L_\text{P}$ and $L_\text{R}$. 
	
	We minimize this constrained objective function above with Mini-Batch Stochastic Gradient Descent (SGD) and control the learning rating using AdaGrad. We keep the negative pair that maximizes the distance with the target user-item pair ($\max F(e_\text{tar}^{(*)},e_\text{ne}^{(*)})$) when we sample negative pairs. Our training process is shown as Algorithm 1.
	
	\begin{algorithm}[!h]
		\caption{MML Training process}
		\begin{algorithmic}[1]
			\REQUIRE User set $U$; item set $I$; user-item rating matrix $R$; margins $mr^u$, $mr^i$, and $mr^l$; hyperparameters $\alpha, \lambda, \theta$, and$ \omega $
			\ENSURE User / Item latent vector set $E_U$ / $E_I$, metric matrix $W^{U}, W^{I}, W^{UI}$.\\
			\STATE Select a batch $B$ with $N$ positive user-item pairs. \\
			\FORALL {$B\in U,I$}
			\FORALL {user-item positive pair ($a,c$)}
			\STATE Sample 2 negative user-item ($a,d$), ($b,c$) pairs to build two triplets.
			\STATE Calculate $L_\text{MML}^\text{EX}$ with Eq.(\ref{eq7}).
			\STATE For $a$, sample 1 similar user $f$ and 1 dissimilar user $g$ with Eq.(\ref{eq8}). Also sample a similar item and a dissimilar item for $c$.
			\STATE Calculate $L_\text{MML}^\text{LA}$ with Eq.(\ref{eq9}).
			\STATE Calculate $L_\text{MMA}$ across batch $B$.
			\ENDFOR
			\WHILE {not converge}
			\STATE Calculate gradients.
			\STATE Update $p_u$ and $q_i$ with AdaGrad on Eq.(\ref{eq12}).
			\STATE Update $W^U$, $W^I$ and $W^{UI}$ with AdaGrad on Eq.(\ref{eq12}).
			\STATE Update $mr^u$, $mr^i$ and $mr^l$ with AdaGrad on Eq.(\ref{eq12}).
			\ENDWHILE
			\ENDFOR
			\RETURN User / Item latent vector set $E_U$ / $E_I$; metric matrix $W^{U}, W^{I}, W^{UI}$.
		\end{algorithmic}
	\end{algorithm}
	\subsection{Comparison with Collaborative Metric Learning}
\begin{figure}[htbp]
		\centering
		\includegraphics[width=0.8\columnwidth]{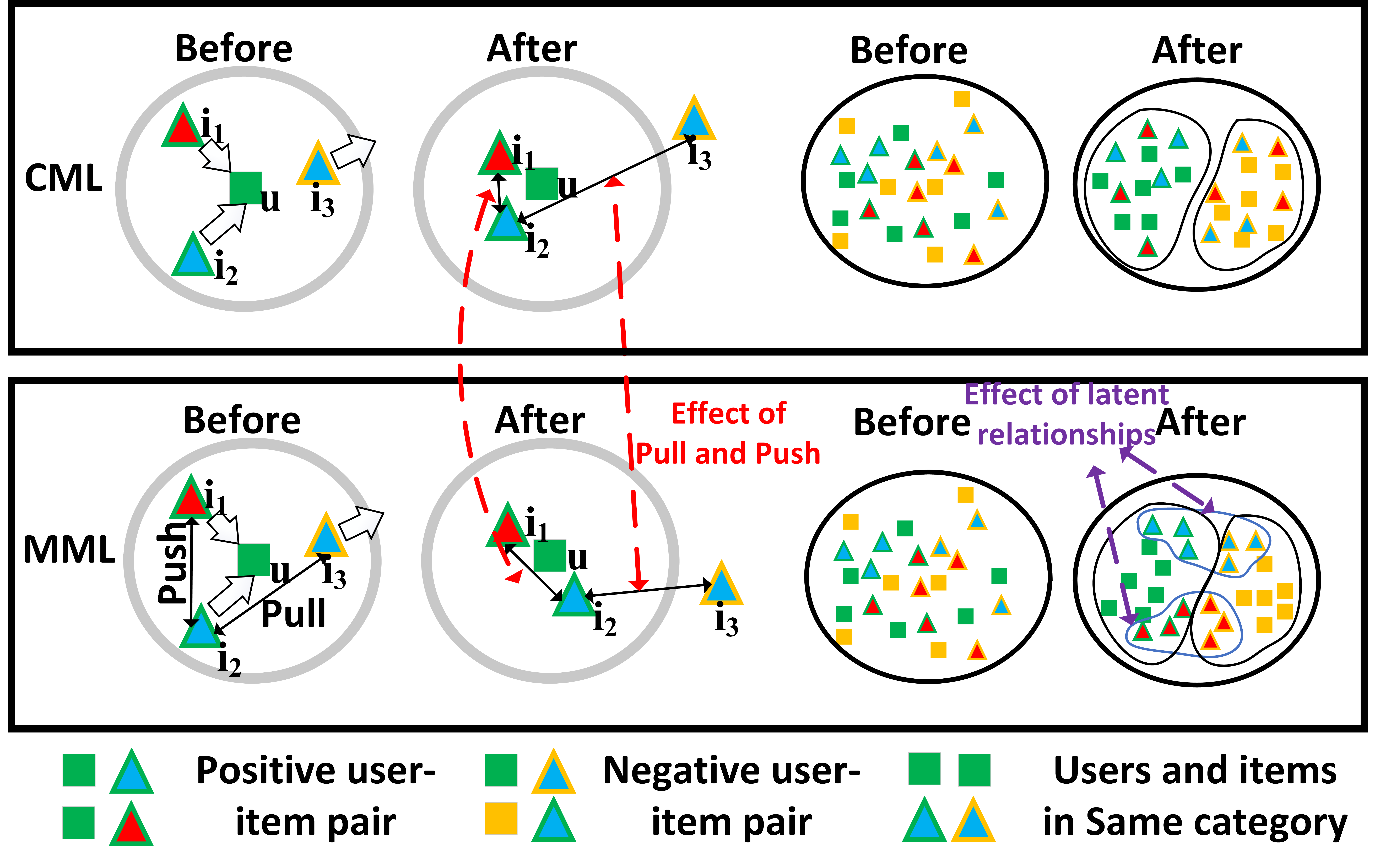}
		\caption{Comparison between CML and MML. The red dashed line shows the effect of Pull and Push. The purple dashed line shows the effect of consideration of latent relationships.}
		\label{Fig:MME3}
	\end{figure}

	We compare our proposed model with a representative model, Collaborative Metric Learning (CML)  \cite{1hsieh2017collaborative} in detail. MML borrows the idea of metric learning, which is similar to CML. However, our model has essential differences compared with CML (shown in Fig.\ref{Fig:MME3}): 
	
	First, CML utilizes only the user-item pair to build the objective function, which focuses on the explicit relationships in the user-item matrix. As shown in Fig.\ref{Fig:MME3}, CML pulls the items $i_1$,$i_2$ to the user $u$ and pushes away item $i_3$. However, note that there are latent relationships hidden in the user-item matrix. So the items in the same similar-pair set ($i_2,i_3$) should be embedded closer, while $i_1,i_2$ should be embedded with a longer distance. MML considers this situation, using Push and Pull for both explicit and latent relationships, to achieve more accurate and unbiased embedding results. 
	
	Second, as shown in the right part of Fig.\ref{Fig:MME3}, MML can learn a direct and visible embedding result because of the latent relationship formulation $L^\text{LA}_\text{MML}$. In $L^\text{LA}_\text{MML}$, MML considers the relationships between same categories (users or items). So MML is able to gather the entities of the same category closer than CML, which is also a great improvement on explainability.
	
	Moreover, CML directly employs Euclidean distance to measure the relationships between different users and items, ignoring the importance variety of different vectors. While MML is able to learn a more accurate metric $W$ for users, items, and user-item respectively, which fine-grained measures the relationships. CML utilizes the fixed margin for all entities, while MML considers the different criteria for different users and items, and employs the adaptive margins to add personality to our model. 
	
	Finally, the objective function of MML with a uniform format ($L^\text{EX}_\text{MML}, L^\text{LA}_\text{MML}$) does not distinguish users and items like CML, which is more feasible and effective. MML employs two regularizations to relieve overfitting situations, especially when the dataset is sparse and unbalanced.
	
	\section{Evaluation} 
	
	In this section, we first describe the experimental settings,
	including datasets, baselines, parameter setting, and implementation details. Subsequently, we conduct extensive experiments to answer the following research questions:
	
	\textbf{RQ1}: How is the effectiveness of MML? Can it provide a competitive performance compared with baselines on the matrix embedding task at a proper running time? \textbf{RQ2}: How do the hyperparameters affect the performance of MML?Which are the optimal values? \textbf{RQ3}: How does the proposed model benefit the neural-network-based recommendation models with Top-K recommendation? \textbf{RQ4}: How do the learned metric benefit the matrix embedding and recommendations? What is the effectiveness of regularization to avoid overfitting? \textbf{RQ5}: What is the embedding performance of MML on million-scale dataset? What is the comparison between MML and other SOTA ME models?
	
	\subsection{Experimental Settings}
	\subsubsection{Datasets}
	
	We conduct experiments on datasets from Amazon.com\footnote[1]{https://jmcauley.ucsd.edu/data/amazon} (we use Amazon as the abbreviation of the Sports and Outdoors dataset in Amazon in this paper) and Yelp for RecSys\footnote[2]{https://www.kaggle.com/c/yelp-recsys-2013}. Moreover, we collect two datasets from Taobao\footnote[3]{https://www.taobao.com} and Jingdong\footnote[4]{https://re.jd.com/} as supplementary to validate our method \cite{105DBLP:journals/nn/XuYHWZYX19,130DBLP:journals/tkdd/XuYWHZYX20}. All the datasets contain ratings ranging from 1 to 5. We divide the datasets: 60\% as the training set, 20\% as the test
	set and 20\% as the validation set with 5-cross-validation, and treat ratings less than 3 as negative samples for recommendations \cite{111DBLP:conf/aaai/LiZZQZHH20,105DBLP:journals/nn/XuYHWZYX19}. Table \ref{Tab1} summarizes the details of datasets.
	\begin{table}[htbp]
		\centering
		\caption{The Datasets' Characteristics.}
		\label{Tab1}
		\resizebox{0.4\textwidth}{!}{
			\begin{tabular}{lcccc}
				\hline\hline
				Dataset     & Amazon  & Yelp & Taobao & Jingdong
				\\ \hline
				\#user    & 30,759  & 45,980 & 10,121 & 8,031
				\\ \hline
				\#item    & 16,515  & 11,537  & 9,892 & 3,025
				\\ \hline
				\#rating    & 285,644  & 229,900 & 49,053 & 25,152
				\\ \hline
				\#item labels    & 36  & 24 & 17 & 12
				\\ \hline
				Sparsity    & 0.051\%  & 0.043\% & 0.049\% & 0.12\%
				\\ \hline\hline
		\end{tabular}}
	\end{table}
	
	\begin{table*}[htbp]
		\centering
		\caption{Relationship measurement and loss function comparison with baselines. $\lambda$ is the hyperparameter, ($u,i$) means a positive pair $r_{ui} \ne 0$, ($u,i^-$) means a negative pair $r_{ui} = 0$, $b_{ui}$ is a learned sharing parameter.}
		\label{Tab21}
		\resizebox{0.8\textwidth}{!}{
			\begin{tabular}{c|cc}
				\hline\hline
				Models     & Relationship measurement  & Loss Function 
				\\ \hline
				WRMF   & $F(u,i) = p_u{q_i^T}$  & $\sum\limits_{u,i} {{{({r_{ui}} - {p_u}q_i^T)}^2} + \lambda (\sum\limits_u {{{\left\| {{p_u}} \right\|}^2}} ) + \lambda (\sum\limits_i {{{\left\| {{q_i}} \right\|}^2}} )} $
				\\ \hline
				CML    & ${F}(u,i) = \left\| {{p_u} - {q_i}} \right\|_{\text{Euc}}^2$  & $\sum\limits_{u,i,{i^ - }} {{{\left| {F(u,i) - F(u,{i^ - }) + m} \right|}_ + }} $  
				\\ \hline
				IML    & ${F}(u,i) = \left\| {{p_u} - {q_i}} \right\|_{\text{Euc}}^2$  & $\sum\limits_{u,i,{i^ - }} {{{\left| {F(u,i) - F(u,{i^ - }) + m} \right|}_ + }} $ 
				\\ \hline
				CRML    & ${F}(u,i) = \left\| {{p_u} - {q_i}} \right\|_{\text{Euc}}^2$  & $\sum\limits_{u,i,{i^ - }} {L({p_u},q{}_i) + L({p_u},{b_{ui}}) + L({q_i},{b_{ui}})}$
				\\ \hline
				SML    & ${F}(u,i) = \left\| {{p_u} - {q_i}} \right\|_{\text{Euc}}^2$  & $\sum\limits_{u,i,{i^ - }} {\left( {{{\left| {F(u,i) - F(u,{i^ - }) + {m_u}} \right|}_ + } + {{\left| {F(u,i) - F(i,{i^ - }) + {m_i}} \right|}_ + }} \right) + \lambda {L_{AM}}} $
				\\ \hline
				MML    & ${F}(u,i) = \left\| {{p_u} - {q_i}} \right\|_{W}^2$  & ${L_\text{MML}} + {\omega _\text{P}}{L_\text{P}}+ {\omega _\text{R}}{L_\text{R}}$ (Eq.(\ref{eq12})) 
				\\ \hline\hline
		\end{tabular}}
	\end{table*}
	\subsubsection{Baselines}
	
	To evaluate our proposed model on matrix embedding task, we compare MML with five representative metric learning models, including:
	
	\textbf{WRMF} \cite{7gu2010collaborative,8gemulla2011large} This implicit MF model utilizes an additional case weight to model unobserved interactions. WRMF can also be treated as a basic matrix factorization embedding model on the user-item matrix with inner products. \textbf{CML} \cite{1hsieh2017collaborative} This representative CF model borrows the idea of metric learning to learn a latent space for users and items. Moreover, it is claimed that CML can outperform most state-of-the-art CF models with the metric-based loss function. \textbf{IML} \cite{6ijcai2018-389} This efficient model applies metric learning to unbalanced data for clustering. IML’s contribution is that it splits data into subsets and accelerates the process. \textbf{CRML} \cite{112DBLP:journals/nn/WuZNC20} This is a metric learning model for collaborative recommendations with co-occurrence embedding regularization. It considers the optimization problem as a multi-task learning problem which includes optimizing a primary task of metric learning and two auxiliary tasks of representation learning. \textbf{SML} \cite{121DBLP:conf/aaai/LiZZQZHH20} This is a metric learning model that symmetrically introduces a positive item-centric metric which maintains closer distance from positive items to users and pushes the negative items away from the positive items at the same time with an adaptive margin. We show the relationship measurement and loss function comparison with baselines in Table \ref{Tab21}.
	
	We combine MML with nine different recommendation models to make a top-k recommendation, including two basic recommendation models, and four neural-network-based recommendation models :
	
	\textbf{UBCF} and \textbf{IBCF} \cite{2hu2008collaborative} compute the similarity (Cosine or Euclidean) between users (UBCF) or items (IBCF), and find the target’s k-nearest neighbors to make Top-K recommendations. \textbf{NCF} \cite{5he2017neural} is a state-of-the-art neural-network-based recommendation model which directly combines the latent vectors as the input of the model. As his work claims, NCF can cover some state-of-the-art CF models. \textbf{2IPS} \cite{117DBLP:conf/www/MaZYYCTHC20} is a typical two-stage off-policy policy gradient method. The proposed method explicitly takes into account the ranking model when training the candidate generation model, which helps improve the performance of the whole system. \textbf{NAIS} \cite{125DBLP:journals/tkde/HeHSLJC18} is an attention network, which is capable of distinguishing which historical items in a user profile are more important for a prediction. \textbf{KTUP} \cite{120DBLP:conf/www/0003W0HC19} jointly learns the model of recommendation and knowledge graph completion. It accounts for various preferences in translating a user to an item, and then jointly trains it with a KG completion model by combining several transfer schemes. \textbf{HERec} \cite{118DBLP:journals/tkde/ShiHZY19} is a heterogeneous network embedding based approach for heterogeneous information network (HIN) based recommendation. To embed HINs, it designs a meta-path based random walk strategy to generate meaningful node sequences for network embedding. \textbf{NGCF} \cite{126DBLP:conf/sigir/Wang0WFC19} exploits the user-item graph structure by propagating embeddings on it. This leads to the expressive modeling of high-order connectivity in user-item graph, effectively injecting the collaborative signal into the embedding process in an explicit manner. \textbf{GraphRec} \cite{127DBLP:conf/www/Fan0LHZTY19} provides a principled approach to jointly capture interactions and opinions in the user-item graph, which coherently models two graphs and heterogeneous strengths.
	
	Some RS baselines are two-stage recommendation models which contain the matrix embedding parts. In this paper, we use ME baselines (WRMF/IML/CRML/SML/MML) to substitute these matrix embedding parts in RS models for testing.
	
	\subsubsection{Parameter Setting and Implementation Details}
	
	The implementation of the comparison methods are from the public codes that the authors provided in their papers or open source project. For MML, we set default margins $mr^u=mr^i=mr^l=0.02$. All latent vectors in dimension $k$ = 32, with random initialization (uniform distributions mean: 0.2, viariance: 0.04). The batch size $B$ is 512. We tune the learning rate {0.01, 0.02, 0.05}. Without special explanations, we set balance weight $\lambda =0.5$, similarity threshold $\theta=0.3$, $\omega=0.03$ and $\alpha=0.7$. All these parameters are determined through cross-validation. 
	
	\subsection{Matrix Embedding Validation (RQ1)}
	In this section, we need to validate whether the models can gather the same items and disperse the different ones. Along with this line, we employ spherical k-means on embedding results, with
	$K$ = 10 and 20 clusters. We use Normalized Mutual Information (NMI) as the protocols:
	\begin{equation}
	\begin{aligned}
	\text{NMI}(L,C) = \frac{{\text{Cor}(L,C)}}{{[\text{H}(L) + \text{H}(C)]/2}},
	\end{aligned}
	\label{eq13} 
	\end{equation}
	where $L$ is the set of labels of items and $C$ is the set of clusters. $\text{Cor}(L,C)$ denotes the sum of mutual information between any label $l$ in any cluster $c$. $\text{H}(L)$ and $\text{H}(C)$ denote the entropy for labels and clusters respectively. This metric evaluates the purity of clustering results from an information-theoretic perspective.
	\begin{table}[htbp]
		\centering
		\caption{Normalized Mutual Information with 10 clusters.}
		\label{Tab2}
		\resizebox{0.4\textwidth}{!}{
			\begin{tabular}{l|cccc}
				\hline\hline
				Model     & Amazon  & Yelp & Taobao & Jingdong
				\\ \hline
				WRMF   & 0.3214  & 0.3013 & 0.4215 & 0.4317
				\\ \hline
				CML    & 0.5310  & 0.5010  & 0.5870 & 0.5711
				\\ \hline
				IML    & 0.5613  & 0.5522 & 0.5830 & 0.6001
				\\ \hline
				CRML    & 0.5673  & 0.5444 & 0.6030 & 0.6111
				\\ \hline
				SML    & 0.5723  & 0.5602 & 0.5933 & 0.6092
				\\ \hline
				\textbf{MML}    & \textbf{0.5831}$^*$  & \textbf{0.5621}$^*$ & \textbf{0.6321}$^*$ & \textbf{0.6134}$^*$
				\\ \hline\hline
		\end{tabular}}
	\end{table}
	
	\begin{table}[htbp]
		\centering
		\caption{Normalized Mutual Information with 20 clusters.}
		\label{Tab3}
		\resizebox{0.4\textwidth}{!}{
			\begin{tabular}{l|cccc}
				\hline\hline
				Model     & Amazon  & Yelp & Taobao & Jingdong
				\\ \hline
				WRMF   & 0.2943  & 0.3001 & 0.3255 & 0.3321
				\\ \hline
				CML    & 0.4732  & 0.4638  & 0.5533 & 0.5612
				\\ \hline
				IML    & 0.4831  & 0.4765 & 0.5545 & 0.5532
				\\ \hline
				CRML    & 0.5023  & 0.5122 & 0.5732 & 0.6001
				\\ \hline
				SML    & 0.5313  & 0.5232 & 0.5644 & 0.6011
				\\ \hline
				\textbf{MML}    & \textbf{0.5433}$^*$  & \textbf{0.5564}$^*$ & \textbf{0.6003}$^*$ & \textbf{0.6112}$^*$
				\\ \hline\hline
		\end{tabular}}
	\end{table}
	
	From the NMI evaluation results in Table \ref{Tab2} and Table \ref{Tab3}, we can see that MML outperforms all the baselines for all clustering value $K$ in all four datasets. This result shows two advantages of MML: First, five models with metric learning are much better than traditional model WRMF, which means that metric-based models are more proper for matrix embedding than inner products. Second, MML tackles both explicit and latent relationships and learns a weighed metric matrix, which leads to a more stable performance than CML, IML, CRML, and SML. Note that in Jingdong with 20 clusters, CRML, SML, and MML's performance are very close. But in Amazon and Yelp, MML outperforms both the state-of-the-art baselines. This indicates the advantage of MML in tackling sparse data.  
	
	\begin{figure}[htbp]
		\centering
		\includegraphics[width=0.9\columnwidth]{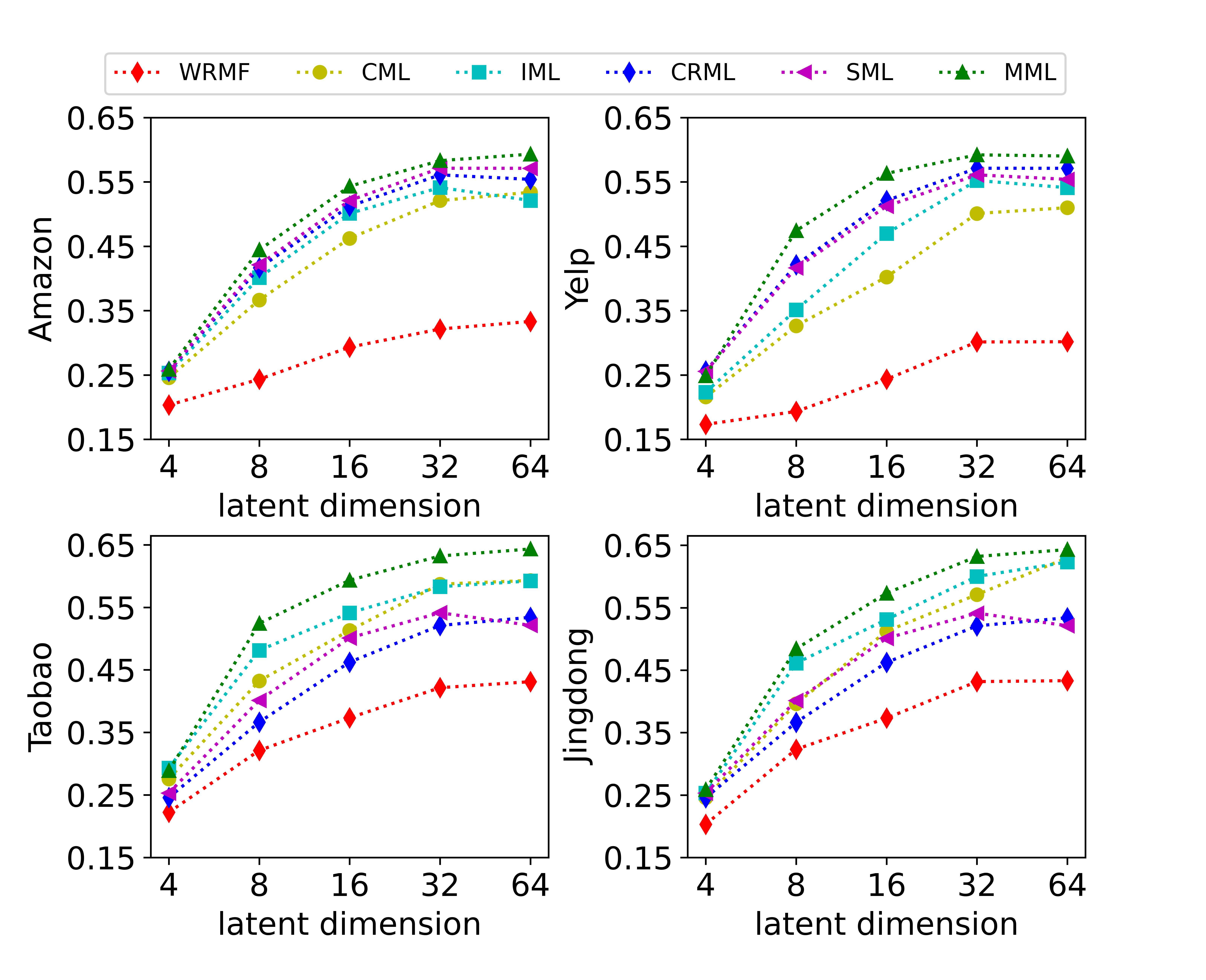}
		\caption{Dimension effect on NMI with 10 clusters.}
		\label{Fig:MME4}
	\end{figure}
	
	Besides, we exploit the effect of latent vector space dimension $k$ (4, 8, 16, 32, and 64) on NMI (Fig.\ref{Fig:MME4}). We notice that almost all ME models' performance is better with high-dimension latent vector space, and MML achieves the best results. High-dimensional data space has a strong representative ability to catch more hidden knowledge of users and items, which can enhance the performance of matrix embedding. Hence, the performance increases fast from 4 to 16. However, note that the increase becomes slower from 16 to 64, which shows the bottleneck of the dimension profit. Note that WRMF achieves the worst performance among baselines, which indicates that in the high dimension latent space, using metric learning is better than inner products in the matrix embedding task.
	\begin{table}[htbp]
		\centering
		\caption{Running time for training process (time unit).}
		\label{Tab4}
		\resizebox{0.4\textwidth}{!}{
			\begin{tabular}{l|cccc}
				\hline\hline
				Time/Epoch     & Amazon  & Yelp & Taobao & Jingdong
				\\ \hline
				CML    & 103$^*$  & 124$^*$  & 68$^*$ & 74$^*$
				\\ \hline
				IML    & 349  & 402 & 156 & 147
				\\ \hline
				CRML    & 112  & 133 & 79 & 89
				\\ \hline
				SML    & 113  & 150 & 88 & 93
				\\ \hline
				MML    & \textbf{110}  & \textbf{130} & \textbf{75} & \textbf{83}
				\\ \hline
				\textit{Ours vs Best}    & \textbf{+7}  & \textbf{+6} & \textbf{+7} & \textbf{+9}
				\\ \hline\hline
		\end{tabular}}
	\end{table}
	
	At last, we also compare the computing time among five metric learning models (Table \ref{Tab4}). CML takes the shortest time each epoch and IML takes the longest. CML's loss function is simple to calculate, so it achieves the best running time. While IML utilizes an iteration metric learning, which means in one epoch, IML learns metric repeatedly on different subsets. Note that two state-of-the-art models, CRML and SML use more time than our proposed model MML. SML combines two different styles of the loss function with two regularizations and three sub loss functions, which adds computation complexity. CRML and MML utilize the same formulation of loss function for explicit and latent relationship embedding, which is easy to compute derivation and speeds up the model's optimization. 
	
	\subsection{Exploring Effect of Hyperparameters (RQ2)}
	
	In this section, we explore the effect of hyperparameters in MML. MML introduces four additional hyperparameters $\alpha, \lambda, \theta$, and $ \omega $. $\alpha \in (0,1)$ controls the learning of explicit and latent relationships. $\lambda \in (0,1)$ controls the learning of EMDL. $\theta \in (0,1)$ restricts the similar-pair set building in latent relationships formulation. $ \omega $ controls the regularizations, which we discuss in the following sections (RQ4). Here we show how the three hyperparameters impact the performance and also shed light on how to set them. We only
	show the results on Amazon and Taobao due to the limitation of space. We use Hitting Ratio (HR) on Top-10 and Top-50 to explore the hyperparameters. We vary one parameter while fixing others as experimental settings.
	
\begin{figure}[htbp]
		\centering
		\includegraphics[width=0.9\columnwidth]{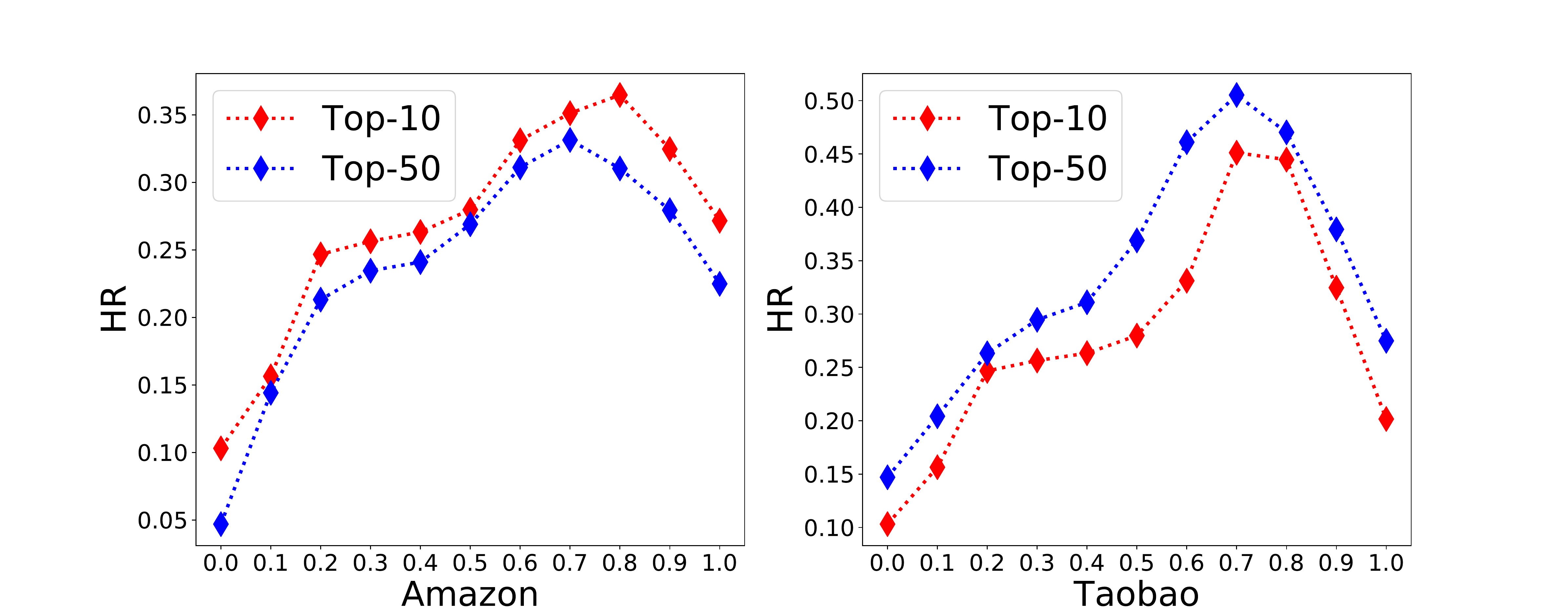}
		\caption{Performance of MML with respect to different values of $\alpha$.}
		\label{Fig:MME12}
	\end{figure}

	As shown in Fig.\ref{Fig:MME12}, the optimal value of $\alpha$ is around 0.7 for both two datasets. And we also observe that the performance improves before $\alpha$ reaches 0.7, then it decreases sharply. Thus the too large value of $\alpha$ will ruin the learning process of metric learning. So we set $\alpha$ to 0.7.
	
\begin{figure}[htbp]
		\centering
		\includegraphics[width=0.9\columnwidth]{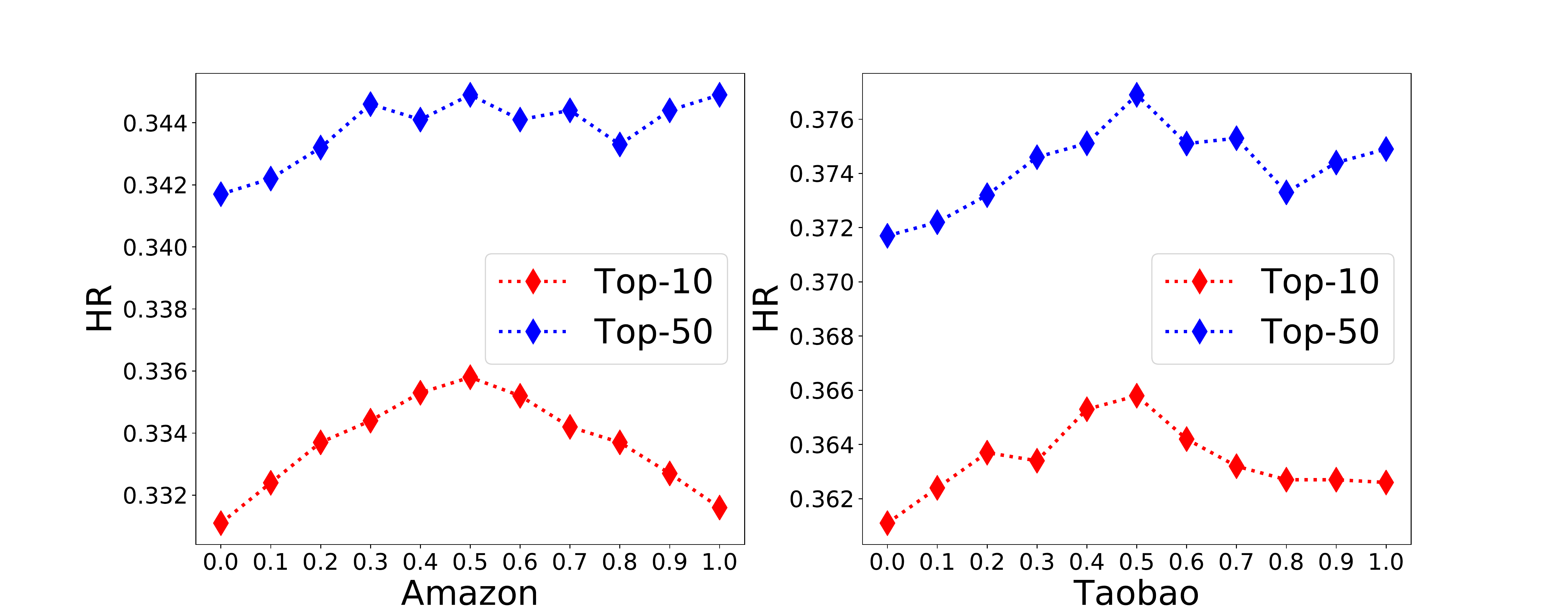}
		\caption{Performance of MML with respect to different values of $\lambda$.}
		\label{Fig:MME13}
	\end{figure}

	As shown in Fig.\ref{Fig:MME13}, the optimal value of $\lambda$ is around 0.5 for both two datasets. When $\lambda$=0.5. it treats the users and items as the same category, which satisfies the assumption of our model (to map users and items into a unified latent space). So we set $\lambda$ to 0.5.
	
\begin{figure}[htbp]
		\centering
		\includegraphics[width=0.9\columnwidth]{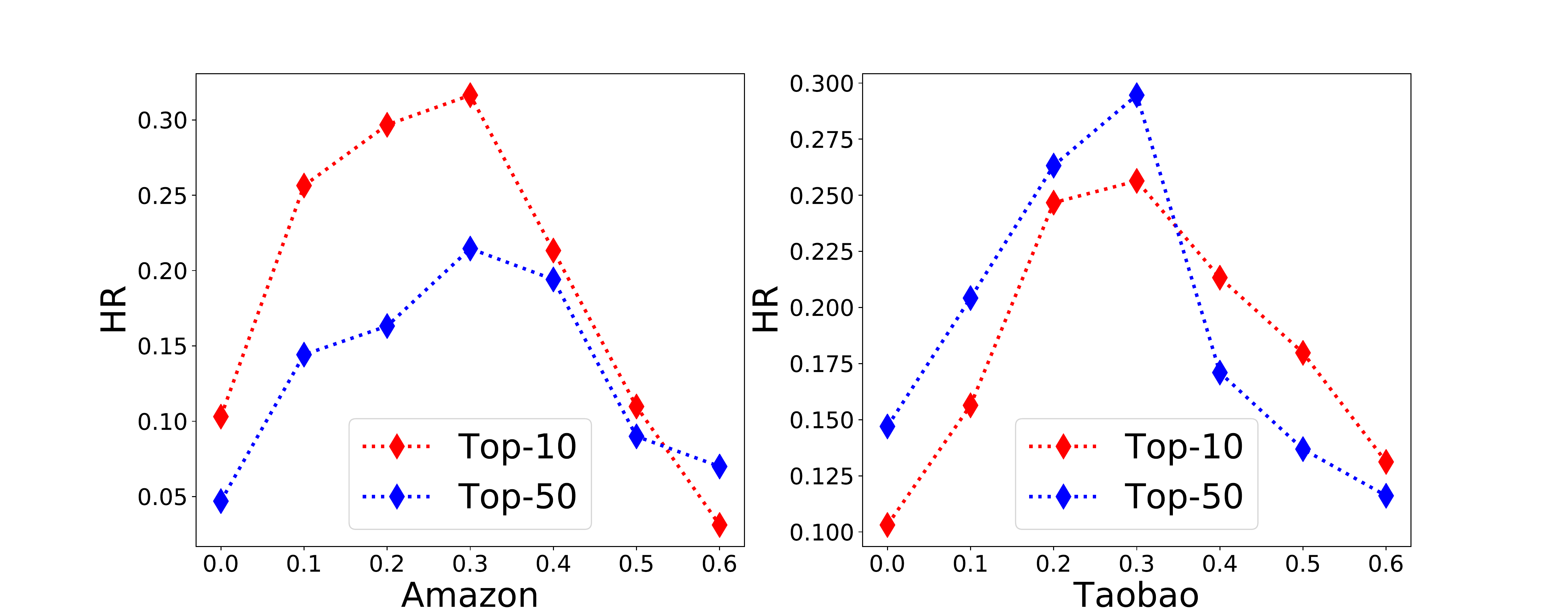}
		\caption{Performance of MML with respect to different values of $\theta$.}
		\label{Fig:MME14}
	\end{figure}

	As shown in Fig.\ref{Fig:MME14}, the optimal value of $\theta$ is around 0.3 for both two datasets. Note that  When $\theta$ is too small, MML behaves minor improvements, which shows there are redundant pairs in similar-pair sets, which hurts the performance. Moreover, if $\theta$ is too large, the performance drops dramatically. So we set $\theta$ to 0.3. 
	
	\vspace{-20pt}
	\vspace{-20pt}
	
	\subsection{Recommendation Validation (RQ3)}
	\begin{table*}[htbp]
		\centering
		\caption{Improvement of recommendation models with different matrix embedding models. $^*$ marks the best performance among baselines.}
		\resizebox{0.7\textwidth}{!}{
			\begin{tabular}{c|c|c|c|c|c|c|c|c|c}
				\hline\hline
				\multirow{2}{1cm}{RS model}&\multirow{2}{1cm}{ME model}&\multicolumn{1}{c}{Amazon}& &\multicolumn{1}{c}{Yelp}&& \multicolumn{1}{c}{Taobao} &&\multicolumn{1}{c}{Jingdong}& \\
				\cline{3-10}
				&      & \textit{HR@50} & \textit{Recall@50} & \textit{HR@50} & \textit{Recall@50} & \textit{HR@50} & \textit{Recall@50} & \textit{HR@50} & \textit{Recall@50} \\
				\hline
				\multirow{5}{1cm}{UBCF} & WRMF    & 0.1372      &  0.2112     &   0.1544    &  0.2002      & 0.3721      &  0.3235     &   0.3313    &  0.3143 \\
				\cline{2-10}
				& CML   & 0.2112      &  0.2411     &  0.2339     &  0.2348     &  0.3826     &  0.4057     &  0.4118     & 0.4756$^*$  \\
				\cline{2-10}
				& IML   & 0.2211      &  0.2333    &   0.2213    &  0.2453      &  0.4052     & 0.4361     &  0.4312     & 0.4123  \\
				\cline{2-10}		
				& CRML   & 0.2312      &  0.2520     &  0.2439     &  0.2548     &  0.4336     &  0.4557     &  0.4780$^*$     & 0.4661  \\
				\cline{2-10}
				& SML   & 0.2328$^*$      &  0.2621$^*$    &   0.2533$^*$    &  0.2653$^*$      &  0.4432$^*$     & 0.4732$^*$     &  0.4610     & 0.4711  \\
				\cline{2-10}		
				& \textbf{MML}   & \textbf{0.2618}      &  \textbf{0.2811}   &   \textbf{0.2794}    &  \textbf{0.2860}      &  \textbf{0.4653}     & \textbf{0.4979}     &  \textbf{0.5167}     & \textbf{0.4790}  \\
				\cline{2-10}
				& \textit{Ours vs Best}   & \textbf{\textit{+12.4\% }}     &  \textbf{\textit{+7.2\%}}    &   \textbf{\textit{+10.3\%} }   &  \textbf{\textit{+7.8\%}}      &  \textbf{\textit{+4.9\%}}     & \textbf{\textit{+5.2\%}}     &  \textbf{\textit{+7.9\% }}    & \textbf{\textit{+1.6\%}}  \\\hline

				\multirow{5}{1cm}{IBCF} & WRMF    & 0.1433      &  0.2011     &   0.1411    &  0.2100      & 0.3543      &  0.3421     &   0.3442    &  0.3301 \\
				\cline{2-10}
				& CML   & 0.1634      &  0.1623     &  0.1777     &  0.2012     &  0.2972     &  0.3022     &  0.3310     & 0.3294  \\
				\cline{2-10}
				& IML   & 0.2011     &  0.2111    &   0.2214    &  0.2433      &  0.3911     & 0.3203     &  0.3882     & 0.4023  \\
				\cline{2-10}
				& CRML   & 0.1934      &  0.1831     &  0.1823     &  0.2213     &  0.3672     &  0.3723     &  0.3890     & 0.3684  \\
				\cline{2-10}
				& SML   & 0.2281$^*$     &  0.2621$^*$    &   0.2710$^*$    &  0.2533$^*$      &  0.4102$^*$     & 0.4302$^*$     &  0.4082$^*$     & 0.4323$^*$  \\
				\cline{2-10}
				& \textbf{MML}   & \textbf{0.3133}      &  \textbf{0.3374}   &   \textbf{0.3332}    &  \textbf{0.3411}      &  \textbf{0.4833}     & \textbf{0.5379}     &  \textbf{0.5100}     & \textbf{0.4990}  \\
				\cline{2-10}
				& \textit{Ours vs Best}   & \textbf{\textit{+37.8\% }}     &  \textbf{\textit{+28.7\%}}    &   \textbf{\textit{+22.9\%} }   &  \textbf{\textit{+34.6\%}}      &  \textbf{\textit{+17.8\%}}     & \textbf{\textit{+25.0\%}}     &  \textbf{\textit{+24.9\% }}    & \textbf{\textit{+15.4\%}}  \\
				\hline

				\multirow{5}{1cm}{NCF} & WRMF    & 0.2041      &  0.2210     &   0.1331    &  0.1994     & 0.3217      &  0.3433     &   0.3614    &  0.3710 \\
				\cline{2-10}
				& CML   & 0.2213      &  0.2561     &  0.2613     &  0.2600     &  0.4231     &  0.4313     &  0.4714     & 0.4705  \\
				\cline{2-10}
				& IML   & 0.2528      &  0.2722    &   0.2810$^*$    &  0.2518      &  0.4303     & 0.4862     &  0.4660     & 0.4913  \\
				\cline{2-10}
				& CRML   & 0.2543      &  0.2771$^*$     &  0.2653     &  0.2693$^*$     &  0.4557$^*$     &  0.4673     &  0.4884$^*$     & 0.4745  \\
				\cline{2-10}
				& SML   & 0.2548$^*$      &  0.2762    &   0.2810$^*$    &  0.2688      &  0.4553     & 0.4879$^*$     &  0.4767     & 0.4933$^*$  \\
				\cline{2-10}
				& \textbf{MML}   & \textbf{0.3318}      &  \textbf{0.3641}   &   \textbf{0.3700}    &  \textbf{0.3660}      &  \textbf{0.5053}     & \textbf{0.5379}     &  \textbf{0.5288}     & \textbf{0.5034}  \\
				\cline{2-10}
				& \textit{Ours vs Best}   & \textbf{\textit{+30.2\% }}     &  \textbf{\textit{+31.3\%}}    &   \textbf{\textit{+31.6\%} }   &  \textbf{\textit{+35.9\%}}      &  \textbf{\textit{+10.8\%}}     & \textbf{\textit{+10.2\%}}     &  \textbf{\textit{+8.2\% }}    & \textbf{\textit{+2.1\%}}  \\
				\hline
				
				\multirow{5}{1cm}{2IPS} & WRMF    & 0.1137      &  0.1040     &   0.1041    &  0.1144     & 0.3091      &  0.2910     &   0.2906    &  0.2959 \\
				\cline{2-10}
				& CML   & 0.2220      &  0.3053     &  0.2958     &  0.2988     &  0.4151     &  0.4115     &  0.4251     & 0.4184  \\
				\cline{2-10}
				& IML   & 0.2234      &  0.3245    &   0.3110    &  0.2764      &  0.4312     & 0.4319     &  0.4555     & 0.4616  \\
				\cline{2-10}
				& CRML   & 0.2411      &  0.3400$^*$     &  0.3253     &  0.3021     &  0.4617$^*$     &  0.4714     &  0.4800     & 0.4645  \\
				\cline{2-10}
				& SML   & 0.2448$^*$      &  0.3312    &   0.3311$^*$    &  0.3452$^*$     &  0.4613     & 0.4867$^*$     &  0.4867$^*$     & 0.4713$^*$  \\
				\cline{2-10}
				& \textbf{MML}   & \textbf{0.2918}      &  \textbf{0.3440}   &   \textbf{0.3706}    &  \textbf{0.3650}      &  \textbf{0.4813}     & \textbf{0.5117}     &  \textbf{0.5012}     & \textbf{0.5023}  \\
				\cline{2-10}
				& \textit{Ours vs Best}   & \textbf{\textit{+16.1\% }}     &  \textbf{\textit{+1.1\%}}    &   \textbf{\textit{+10.6\%} }   &  \textbf{\textit{+5.4\%}}      &  \textbf{\textit{+4.0\%}}     & \textbf{\textit{+4.7\%}}     &  \textbf{\textit{+2.8\% }}    & \textbf{\textit{+6.1\%}}  \\
				\hline
				
				\multirow{5}{1cm}{NAIS} & WRMF    & 0.1184      & 0. 1194    &   0.1172    &  0.1193     & 0.2412      &  0.3329     &   0.3001   &  0.3200 \\
				\cline{2-10}
				& CML   & 0.2313      &  0.2910     &  0.3111     &  0.3200     &  0.4417     &  0.4564     &  0.4428     & 0.4511  \\
				\cline{2-10}
				& IML   & 0.2601      &  0.2813    &   0.3221   &  0.3226      &  0.4754     & 0.4719     &  0.4816     & 0.4776  \\
				\cline{2-10}
				& CRML   & 0.2799      &  0.2997$^*$     &  0.3399     &  0.3411     &  0.4888     &  0.4814     &  0.4904     & 0.4883  \\
				\cline{2-10}
				& SML   & 0.2900$^*$      &  0.2911    &   0.3466$^*$    &  0.3551$^*$      &  0.5012$^*$     & 0.5003$^*$     &  0.5019$^*$     & 0.4933$^*$  \\
				\cline{2-10}
				& \textbf{MML}   & \textbf{0.3111}      &  \textbf{0.3532}   &   \textbf{0.3611}    &  \textbf{0.3588}      &  \textbf{0.5378}     & \textbf{0.5400}     &  \textbf{0.5510}     & \textbf{0.5410}  \\
				\cline{2-10}
				& \textit{Ours vs Best}   & \textbf{\textit{+7.2\%} }      &  \textbf{\textit{+17.8\%} }     &   \textbf{\textit{+5.6\%} }   &  \textbf{\textit{+1.0\%}}      &  \textbf{\textit{+7.3\%}}     & \textbf{\textit{+7.9\%}}     &  \textbf{\textit{+9.7\% }}    & \textbf{\textit{+9.6\%}}  \\
				\hline
				
				\multirow{5}{1cm}{NGCF} & WRMF    & 0.1201      &  0.1209     &   0.1222    &  0.3102     & 0.3000      &  0.3222     &   0.3015    &  0.3132 \\
				\cline{2-10}
				& CML   & 0.2440      &  0.2411     &  0.2946     &  0.2945     &  0.4003     &  0.4013     &  0.4112     & 0.4113  \\
				\cline{2-10}
				& IML   & 0.2531      &  0.2664   &   0.2677    &  0.3011      &  0.5001     & 0.5023     &  0.4954     & 0.4333  \\
				\cline{2-10}
				& CRML   & 0.2679      &  0.2649     &  0.3216     &  0.3364     &  0.5013     &  0.4964     &  0.4755     & 0.5014  \\
				\cline{2-10}
				& SML   & 0.2974$^*$      &  0.2874$^*$    &   0.3454$^*$    &  0.3461$^*$      &  0.5105$^*$     & 0.5009$^*$    &  0.4969$^*$     & 0.5110$^*$  \\
				\cline{2-10}
				& \textbf{MML}   & \textbf{0.3221}      &  \textbf{0.3600}   &   \textbf{0.3646}    &  \textbf{0.3654}      &  \textbf{0.5394}     & \textbf{0.5475}     &  \textbf{0.5564}     & \textbf{0.5433}  \\
				\cline{2-10}
				& \textit{Ours vs Best}   & \textbf{\textit{+8.3\%} }     &  \textbf{\textit{+25.2\%} }    &   \textbf{\textit{+5.5\%} }   &  \textbf{\textit{+5.5\%}}      &  \textbf{\textit{+5.6\%}}     & \textbf{\textit{+9.3\%}}     &  \textbf{\textit{+11.9\% }}    & \textbf{\textit{+6.3\%}}  \\
				\hline
				
				\multirow{5}{1cm}{KTUP} & WRMF    & -      &  -     &   0.1002    &  0.1083     & 0.2842      &  0.3178     &   0.2936    &  0.3132 \\
				\cline{2-10}
				& CML   & -      &  -     &  0.2583     &  0.2584     &  0.4635     &  0.4753     &  0.4500     & 0.4347  \\
				\cline{2-10}
				& IML   & -      &  -    &   0.2677    &  0.2711      &  0.4853     & 0.4879$^*$     &  0.4700     & 0.4613  \\
				\cline{2-10}
				& CRML   & -      &  -     &  0.2813     &  0.2693     &  0.4777     &  0.4773     &  0.4801$^*$     & 0.4645$^*$  \\
				\cline{2-10}
				& SML   & -      &  -    &   0.2817$^*$    &  0.2788$^*$      &  0.4892$^*$     & 0.4879$^*$     &  0.4767     & 0.4633  \\
				\cline{2-10}
				& \textbf{MML}   & -      &  -   &   \textbf{0.3411}    &  \textbf{0.3510}      &  \textbf{0.5211}     & \textbf{0.5321}     &  \textbf{0.5388}     & \textbf{0.4910}  \\
				\cline{2-10}
				& \textit{Ours vs Best}   & -     &  -    &   \textbf{\textit{+17.4\%} }   &  \textbf{\textit{+20.6\%}}      &  \textbf{\textit{+6.1\%}}     & \textbf{\textit{+10.2\%}}     &  \textbf{\textit{+10.9\% }}    & \textbf{\textit{+5.3\%}}  \\
				\hline
				
				\multirow{5}{1cm}{HERec} & WRMF    & -      &  -     &   0.1044    &  0.1027     & 0.2950      &  0.2915     &   0.2945    &  0.2888 \\
				\cline{2-10}
				& CML   & -      &  -     &  0.2568     &  0.2526     &  0.4571     &  0.4364     &  0.4204     & 0.4429  \\
				\cline{2-10}
				& IML   & -      &  -    &   0.2671    &  0.2505      &  0.4509     & 0.4284     &  0.4174     & 0.4468  \\
				\cline{2-10}
				& CRML   & -      &  -     &  0.2700     &  0.2713$^*$     &  0.4717$^*$     &  0.4773$^*$     &  0.4814$^*$     & 0.4712  \\
				\cline{2-10}
				& SML   & -      &  -    &   0.2813$^*$    &  0.2698      &  0.4652     & 0.4679     &  0.4712     & 0.4813$^*$  \\
				\cline{2-10}
				& \textbf{MML}   & -      &  -   &   \textbf{0.3542}    &  \textbf{0.3711}      &  \textbf{0.5333}     & \textbf{0.5279}     &  \textbf{0.5408}     & \textbf{0.5112}  \\
				\cline{2-10}
				& \textit{Ours vs Best}   & -     &  -    &   \textbf{\textit{+20.5\%} }   &  \textbf{\textit{+26.8\%}}      &  \textbf{\textit{+11.5\%}}     & \textbf{\textit{+9.5\%}}     &  \textbf{\textit{+10.9\% }}    & \textbf{\textit{+5.8\%}}  \\
				\hline
				
				\multirow{5}{1cm}{GraphRec} & WRMF    & -      &  -     &   0.1112    &  0.1113     & 0.1942      &  0.2188     &   0.2711    &  0.3009 \\
				\cline{2-10}
				& CML   & -      &  -     &  0.2333     &  0.2534     &  0.4112     &  0.4342     &  0.4432     & 0.4232  \\
				\cline{2-10}
				& IML   & -      &  -    &   0.2577    &  0.2600      &  0.4723     & 0.4631    &  0.4564     & 0.4513  \\
				\cline{2-10}
				& CRML   & -      &  -     &  0.3013     &  0.3023$^*$      &  0.4917     &  0.5000     &  0.5101     & 0.5003  \\
				\cline{2-10}
				& SML   & -      &  -    &   0.3117$^*$    &  0.3000      &  0.5011$^*$     & 0.5001$^*$     &  0.5123$^*$     & 0.5188$^*$ \\
				\cline{2-10}
				& \textbf{MML}   & -      &  -   &   \textbf{0.3655}    &  \textbf{0.3659}      &  \textbf{0.5400}     & \textbf{0.5521}     &  \textbf{0.5601}     & \textbf{0.5531}  \\
				\cline{2-10}
				& \textit{Ours vs Best}   & -     &  -    &   \textbf{\textit{+17.2\%} }   &  \textbf{\textit{+21.9\%}}      &  \textbf{\textit{+7.7\%}}     & \textbf{\textit{+10.3\%}}     &  \textbf{\textit{+9.1\% }}    & \textbf{\textit{+6.6\%}}  \\
				\hline\hline
		\end{tabular}}
		\label{Tab5}%
	\end{table*}%
	
	In this section, we validate the quality of embedding on recommendations. We treat six ME models as matrix embedding models, combining with nine popular recommendation models to make a Top-k recommendation. Hitting Ratio (HR) and Recall are employed to evaluate the recommendations. All the results, including our proposed model and baselines, achieve the best performance while keeping all hyperparameters at their optimal settings. The results on four datasets are shown in Table \ref{Tab5}. Note that KTUP, HERec, and GraphRec are knowledge-graph based recommendation models, and the Amazon data does not provide the KGs.
	
	In all datasets, our proposed model MML outperforms all the ME baselines with three recommendation models, which is a noticeable improvement. In detail, WRMF performs the worst, especially when it combines with NCF, 2IPS, KTUP, and HERec. Note that WRMF is the only method that utilizes the inner products as the measurement for relationships. This result proves the effect of metric learning. When we compare UBCF and IBCF with different metric based models, it is interesting that CML's performance drops significantly, even worse than WRMF. The reason is that CML treats users as the center of embedding, which affects the items' embedding. Although IML also utilizes the idea of CML, the computation iteration of IML can make compensation to some extent. However, in our proposed model, we treat items and users as the same category to ensure accuracy. For NCF, because our models take more knowledge (the latent relationships) into consideration than CML and IML, it also improves an average of 20\% over baselines. 
	
	Compared with two state-of-the-art models, CRML and SML, we notice that the improvement is more obvious on Amazon and Yelp than on Taobao and Jindong. Taking deep insight, MML utilizes different relationships, including explicit and latent ones. With these relationships, MML can relieve the data-sparse issue. While SML only considers user-item, item-time relationships. For CRML, it combines two style loss functions, which we argue it damage the embedding performance to some extent. For cooperating with GNN based models (GraphRec), MML could reach the best recommendation performance. At last, the most important factor is that MML learns a weighted metric matrix $W$, and uses $W$ to calculate the distance in recommender systems, which is a significant improvement.
	
	Moreover, we explore the perfromance enhancement for neural-network based recommendation models (basic MLP \cite{122DBLP:conf/recsys/YangBW18}, NCF, 2IPS, KTUP, and HERec) using MML as a preprocessing for Top-10 and Top-50 recommendation. We conduct experiments on Yelp and Taobao. The results are shown in Fig.\ref{Fig:MME15}: 
	
	\begin{figure}[h] \centering  
		\subfigure{ \label{fig:subfig:f1}     
			\includegraphics[width=0.46\columnwidth]{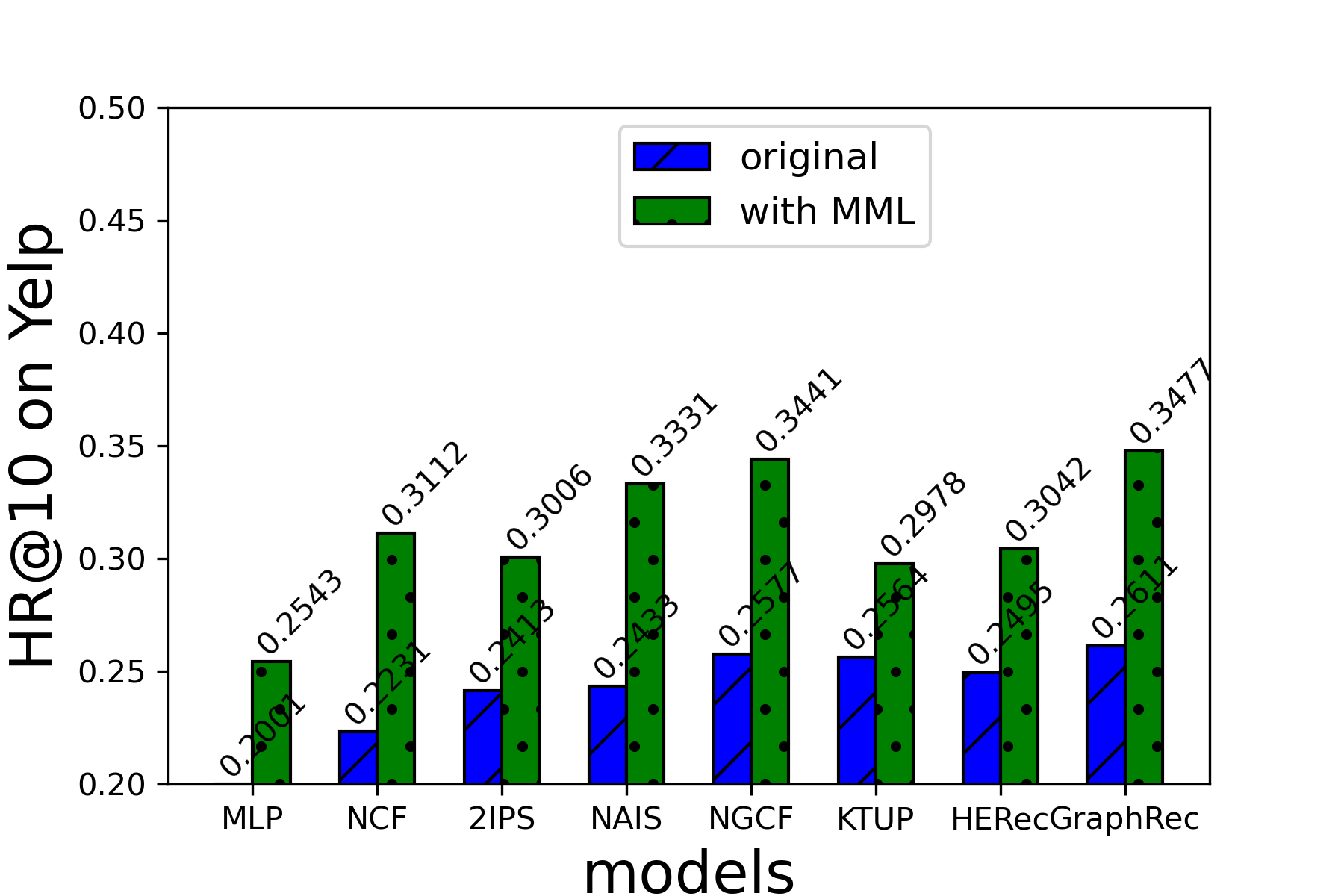}  
		}     
		\subfigure { \label{fig:subfig:f2}     
			\includegraphics[width=0.46\columnwidth]{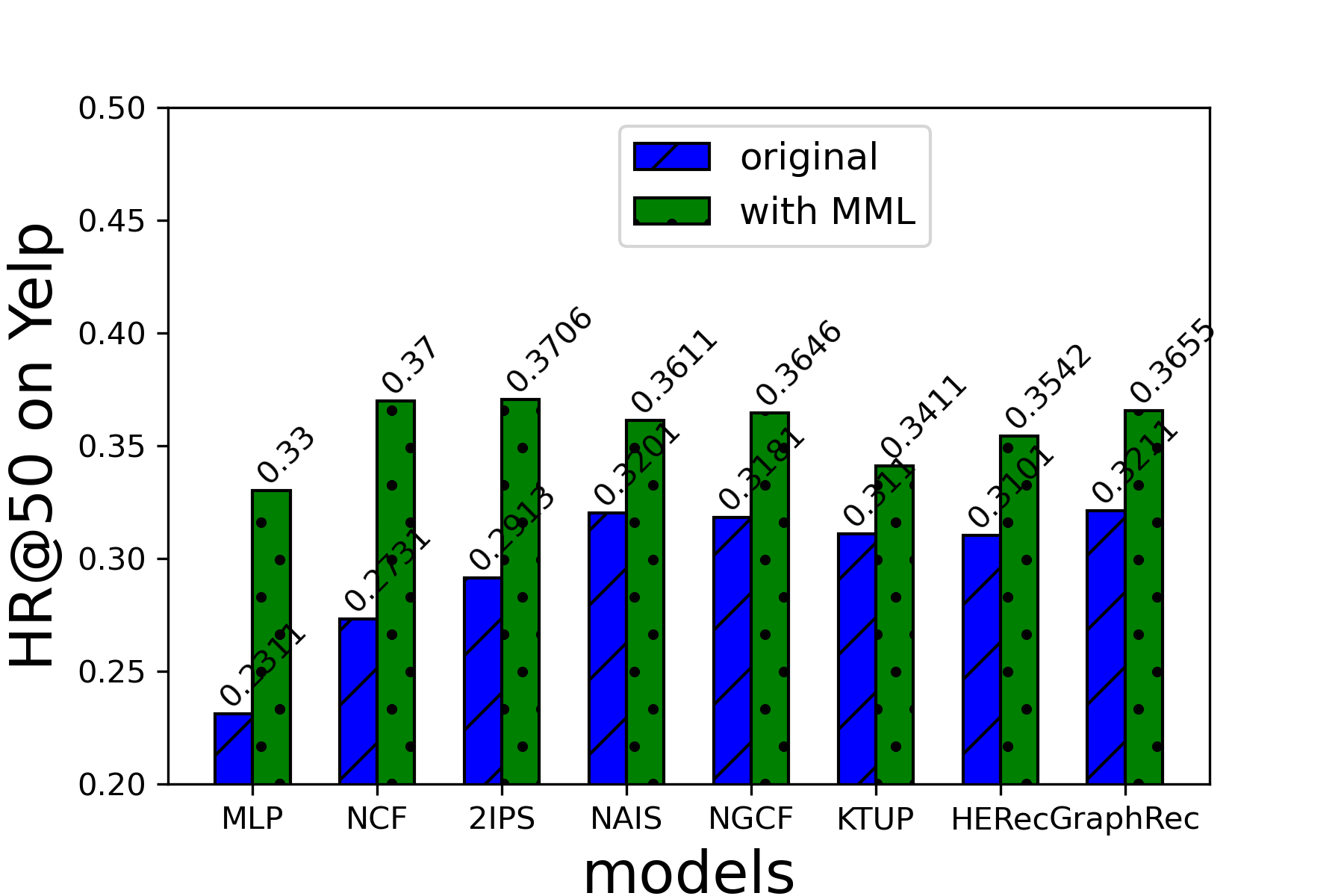}    
		}
		\subfigure { \label{fig:subfig:q1}
			\includegraphics[width=0.46\columnwidth]{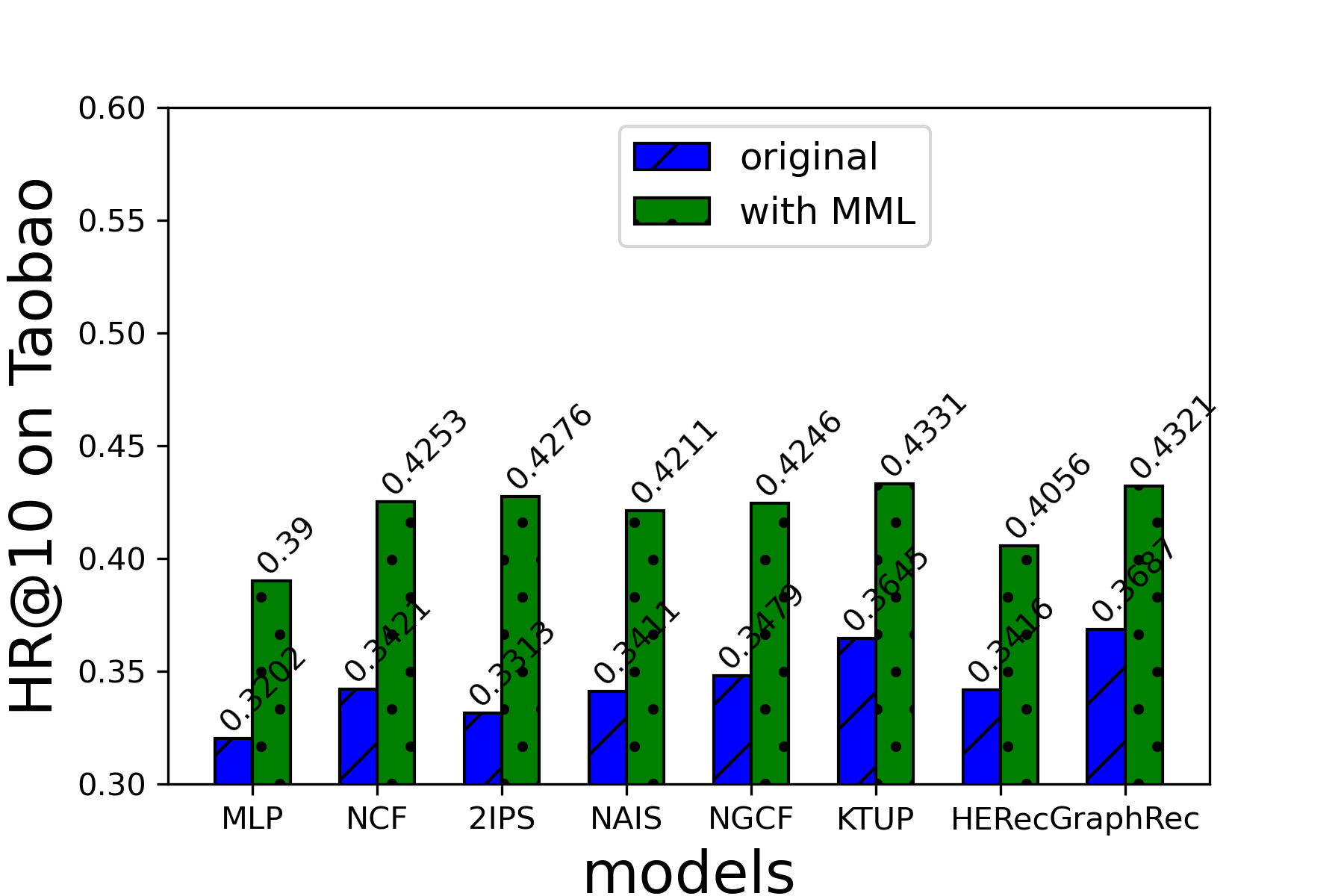} 
		}     
		\subfigure{ \label{fig:subfig:q2}     
			\includegraphics[width=0.46\columnwidth]{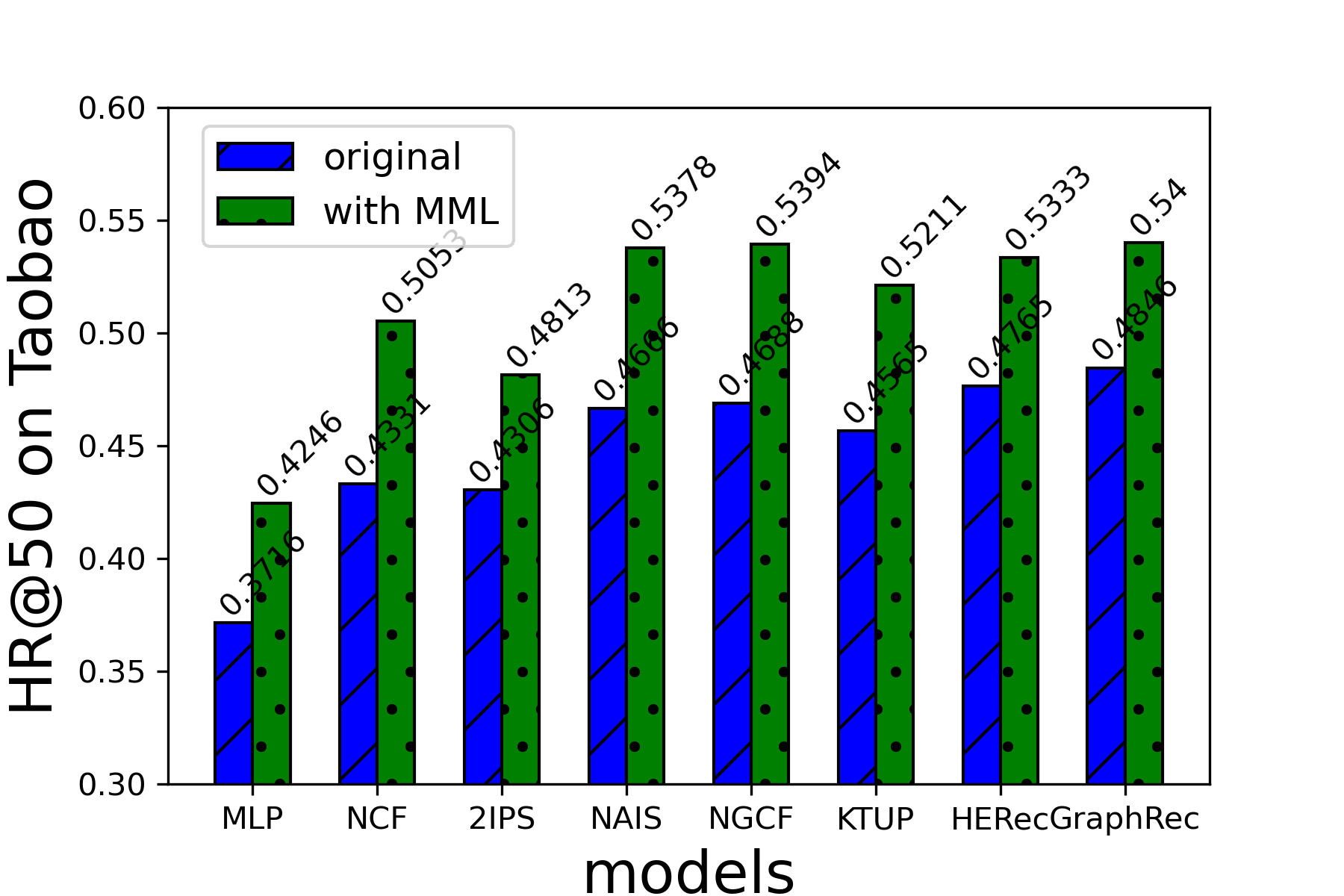}     
		}          
		\caption{Performance Gain with MML for Neural-Network based Recommendation models.}     
		\label{Fig:MME15}     
	\end{figure}
	
	Note that with MML as a preprocessing for neural-network-based recommendation models, HR performance is enhanced over all baselines on both datasets. Specifically, MLP is the basic neural-network-based model that directly inputs latent vectors to predict ratings. The performance gain over MLP indicates the accuracy of latent vectors MML has learned. And for some of the state-of-the-art NN based models, MML can improve the HR performance by average 15\% on Yelp, 17\% on Taobao.
	
	\subsection{Exploring the Effect of MML's Component (RQ4)}
	
	In this section, we explore the effect of learned metric in our proposed model. We separate MML with each component, and rebuild the following models: 
	
	1) EUC-MML: Use ${F^E}(a,b) = \left\| {{e_a} - {e_b}} \right\|_{\text{Euc}}^2$ to replace $F$ in MML (compare Euclidean with Learned metric).
	
	2) W-MML: Use one $W$ to replace $W^{U}$, $W^{I}$ and $W^{UI}$ in MML (compare fixed metric matrix with multi-metric matrix). 
	
	3) M-MML: Use one fix margin $mr$ to replace $mr^u$, $mr^i$ and $mr^l$ (compare fixed margin with adaptive margin). 
	
	4) NP-MML: Use $L_{\text{MML}}$ without restriction $L_{\text{P}}$ ($\omega_P$=0).
	
	5) NR-MML: Use $L_{\text{MML}}$ without restriction $L_{\text{R}}$ ($\omega_R$=0).
	
	\begin{table*}[htbp]
		\centering
		\caption{Effect of each component in MML (including learned metric, adaptive margin and regularizations) on NMI and HR performance.}
		\label{Tab22}
		\resizebox{0.8\textwidth}{!}{
			\begin{tabular}{|c|c|c|c|c|c|c|c|c|c|c|c|}
				\hline\hline
				Models     & Datasets  & \multicolumn{2}{c|}{Performance}  &Models &Datasets & \multicolumn{2}{c|}{Performance}&Models &Datasets & \multicolumn{2}{c|}{Performance}
				\\ \hline
				\multirow{8}{*}{EUC-MML}     & \multirow{2}{1cm}{Amazon}  & NMI & 0.3132 &\multirow{8}{*}{W-MML}     & \multirow{2}{1cm}{Amazon}  & NMI & 0.4212
				&\multirow{8}{*}{M-MML}     & \multirow{2}{1cm}{Amazon}  & NMI & 0.4309 
				\\ \cline{3-4} \cline{7-8} \cline{11-12}
				&& HR@50& 0.2231
				&&& HR@50& 0.2744
				&&& HR@50& 0.2823
				\\  \cline{2-4} \cline{6-8} \cline{10-12}
				&\multirow{2}{1cm}{Yelp}  & NMI & 0.3923&
				&\multirow{2}{1cm}{Yelp}  & NMI & 0.4832&
				&\multirow{2}{1cm}{Yelp}  & NMI & 0.4992
				\\ \cline{3-4} \cline{7-8} \cline{11-12}
				&& HR@50& 0.2702
				&&& HR@50& 0.3212
				&&& HR@50& 0.3212
				\\ \cline{2-4} \cline{6-8} \cline{10-12}
				&\multirow{2}{1cm}{Taobao}  & NMI & 0.4734&
				&\multirow{2}{1cm}{Taobao}  & NMI & 0.5621&
				&\multirow{2}{1cm}{Taobao}  & NMI & 0.5712
				\\ \cline{3-4} \cline{7-8} \cline{11-12}
				&& HR@50& 0.3823
				&&& HR@50& 0.4222
				&&& HR@50& 0.4332
				\\ \cline{2-4} \cline{6-8} \cline{10-12}
				&\multirow{2}{1cm}{Jingdong}  & NMI & 0.5012&
				&\multirow{2}{1cm}{Jingdong}  & NMI & 0.5432&
				&\multirow{2}{1cm}{Jingdong}  & NMI & 0.5543
				\\ \cline{3-4} \cline{7-8} \cline{11-12}
				&& HR@50& 0.4011
				&&& HR@50& 0.4532
				&&& HR@50& 0.4733
				\\ \hline\hline
				
				Models     & Datasets  & \multicolumn{2}{c|}{Performance}  &Models &Datasets & \multicolumn{2}{c|}{Performance}&Models &Datasets & \multicolumn{2}{c|}{Performance}
				\\ \hline
				\multirow{8}{*}{NP-MML}     & \multirow{2}{1cm}{Amazon}  & NMI & 0.5637 &\multirow{8}{*}{NR-MML}     & \multirow{2}{1cm}{Amazon} & NMI & 0.5766
				&\multirow{8}{*}{MML}     & \multirow{2}{1cm}{Amazon}  & NMI & 0.5831
				\\ \cline{3-4} \cline{7-8} \cline{11-12} 
				&& HR@50& 0.3132
				&&& HR@50& 0.3213
				&&& HR@50& 0.3318
				\\ \cline{2-4} \cline{6-8} \cline{10-12}
				&\multirow{2}{1cm}{Yelp}  & NMI &0.5600& 
				&\multirow{2}{1cm}{Yelp}  & NMI &0.5431&
				&\multirow{2}{1cm}{Yelp}  & NMI &0.5621
				\\ \cline{3-4} \cline{7-8} \cline{11-12}
				&& HR@50&  0.3611
				&&& HR@50& 0.3550
				&&& HR@50& 0.3700
				\\ \cline{2-4} \cline{6-8} \cline{10-12}
				&\multirow{2}{1cm}{Taobao}  & NMI & 0.5833&
				&\multirow{2}{1cm}{Taobao}  & NMI & 0.6131& 
				&\multirow{2}{1cm}{Taobao}  & NMI & 0.6321
				\\ \cline{3-4} \cline{7-8} \cline{11-12}
				& & HR@50& 0.4979
				& & & HR@50& 0.5051
				& & & HR@50& 0.5053
				\\ \cline{2-4} \cline{6-8} \cline{10-12}
				&\multirow{2}{1cm}{Jingdong}  & NMI & 0.6014& &\multirow{2}{1cm}{Jingdong}  & NMI & 0.5932 & &\multirow{2}{1cm}{Jingdong}  & NMI & 0.6134
				\\ \cline{3-4} \cline{7-8} \cline{11-12}
				&& HR@50& 0.5098
				&&& HR@50& 0.5132
				&&& HR@50& 0.5288
				\\ \hline
				\hline
		\end{tabular}}
	\end{table*}
	We conduct experiments on four datasets with NMI with 10 clusters and HR@50. The effect of different component in MML is shown in Table \ref{Tab22}.
	
	We notice that MML achieves the best performance (NMI and HR) over all four datasets. Specifically, EUC-MML performs worst than other models, which indicates that in our proposed model, Euclidean is not the proper metric for matrix embedding tasks and recommendations. The simple Euclidean metric may be not suitable for measuring the distance in high-dimension latent space. So metric learning for matrix embedding is necessary for complex NN-based recommendation models to tackle large scale sparse data. According to the comparison between W-MML and MML, it indicates that the metric between users, items, and user-item should be learned respectively to achieve a better result. It is obvious that the features of users and items are different, so learning different $W$ is reasonable. The same explanation can be applied for the comparison between M-MML and MML which indicates the advantage of adaptive margins.  
	
	For NP-MML and NR-MML, we can ensure the effect of regularization. Although the performance of MML is better than NP-MML and NR-MML with a small gap, both regularization can enhance the model by avoiding overfitting.  
	
\begin{figure}[htbp]
		\centering
		\includegraphics[width=1\columnwidth]{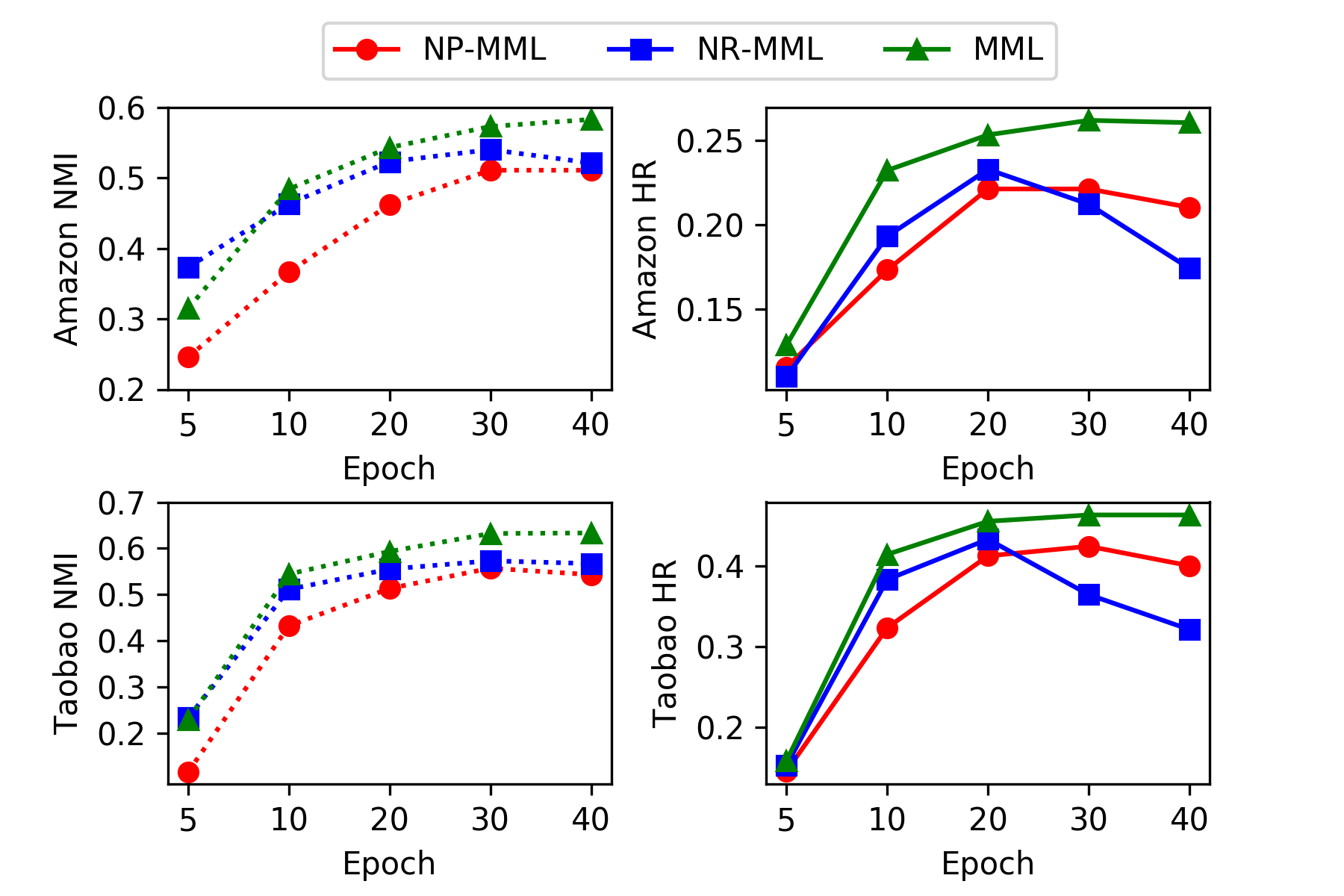}
		\caption{Overfitting analysis on NMI and HR.}
		\label{Fig:MME5}
	\end{figure}

	To evaluate the effect for $L_\text{P}$, $L_\text{R}$ to relieve the overfitting situation, we run NP-MML, NR-MML and MML on Amazon and Taobao to see the performance (NMI with 10 clusters and HR@50) changing with different epochs. The performance changing with epochs is shown in Fig.\ref{Fig:MME5}.
	
	From the results, we can see that NR-MML achieves its best NMI performance within 20 epochs, while NP-MML and MML achieve their best within 30 epochs. Although the best performance of these three models is in the same level, we notice that NR-MML's HR decreases rapidly after 20 epochs, which is a significant overfitting phenomenon. $L_\text{R}$ is to restrict the margins. When the epochs add up, a fixed margin can not measure the detailed distance between high-dimension latent vectors, which leads to the overfitting situation. While $L_\text{P}$ is to restrict the latent vectors. Without $L_\text{P}$, our proposed model suffers the biased and de-centered embedding results. So NR-MML's performance can not be improved after 20 epochs by these biased latent vectors. 
	
	Compared with both models, MML can restrict the embedding results and tune the margins over epochs, with two regularizations $L_\text{R}$ and $L_\text{P}$. Note that MML's performance is stable without sharp fluctuation, which also indicates robustness and effectiveness. Comprehensively, with the consideration of accuracy, efficiency, and overfitting, MML achieves a more stable and feasible performance than all these rebuilt models.
	
	\subsection{Million-scale Embedding Validation (RQ5)}
	
	We conduct WRMF, ConvMF \cite{134DBLP:conf/recsys/KimPOLY16} (a SOTA model which combines GCN with MF for recommendations, widely employed as benchmarks), SML and our proposed MML on a million-scale dataset (Amazon Beauty, with 6,403,006 users, 1,660,119 items, 14,771,988 ratings, with 2.3070 ratings per user and 0.0001\% sparsity) \cite{133DBLP:journals/tnn/HanZXZZYZ20}. Note that this work focuses on the embedding procedure, we only utilize the models' embedding results for validations. Specifically, we use NMI with 20 clusters and running time unit as metrics, as shown in Table \ref{Tab31}:
	
	\begin{table}[tbp]
		\centering
		\caption{Million-scale Performance (with Amazon Beauty dataset).}
		\label{Tab31}
		\resizebox{0.4\textwidth}{!}{
			\begin{tabular}{l|cccc}
				\hline\hline
				Model     & WRMF  & ConvMF & SML & \textbf{MML}
				\\ \hline
				NMI-20   & 0.0832  & 0.2026$^*$ & 0.1813 & \textbf{0.1983}
				\\ \hline
				Time    & 19,331($\pm 43$)  & 21,334($\pm 178$)  & 11,864($\pm 57$) & \textbf{10,333($\pm 62$)}$^*$
				\\ \hline\hline
		\end{tabular}}
	\end{table}

    Note that ConMF performs better on NMI than MML (2.12\%). The reason is that ConvMF enriches the dataset by convolution operations with CNN framework. However, limited by the scale of dataset, ConMF need more running time (almost 100\%) than metric learning-based model (SML and MML) for computing parameters. Considering the trade-off between effectiveness and accuracy, MML achieves a stable performance with acceptable running time on million-scale datasets.
	
	\section{Related works}
	
	\subsection{Matrix Embedding (ME)}
	Matrix Embedding (ME) is usually employed as a pre-procedure for recommender systems, which projects the user-item matrix into latent spaces for users and items \cite{13nilashi2018recommender,100DBLP:journals/tkde/WangTW018}. In general, recommender systems without neural networks always use matrix factorization with inner products to get the users' and items' latent representations in a learned latent vector space \cite{11ricci2015recommender,14he2016interactive}. There are some popular matrix factorization based matrix embedding models, such as WRMF \cite{7gu2010collaborative} and SVD \cite{108DBLP:journals/kais/NikolakopoulosK19}. Matrix factorization with inner products works well with some small datasets like Movielens \cite{15harper2016movielens}. However, because of some limitations that we introduced above, inner products weaken the performance of recommender systems (collaborative filtering models, user or item-based models) in many aspects.
	
	Recently, as a powerful tool of deep learning, the neural network has been widely applied in recommender systems \cite{12zhang2017deep,16xu2018exploiting}. The ability of neural networks enhances the recommender system to the next level. As a preprocessing for recommender systems, traditional matrix embedding models can be enhanced by neural networks. \cite{106DBLP:journals/tnn/FanW17} develops the neural network framework for MF, and proposes a neural-network based MF model.  However, because of weak interpretability \cite{17senthilkumaran2009image} and the strong fitting ability for neural networks, most researchers focus on the neural networks' framework rather than the quality of the matrix embedding. Checking the existing recommendation models \cite{106DBLP:journals/tnn/FanW17,108DBLP:journals/kais/NikolakopoulosK19,120DBLP:conf/www/0003W0HC19,17senthilkumaran2009image,132DBLP:conf/kdd/KabburNK13}, they usually treat the embedding results as a default and limit the explanation for matrix embedding in details, like LightGCN \cite{131DBLP:conf/sigir/0001DWLZ020}. While in this paper, we argue that as important representative vectors for users and items, matrix embedding models do affect the performance of recommendations and should be more focused.
	
	\subsection{Metric Learning (ML)}
	Metric learning (ML) is a research spot for image recognition, clustering, and recommendation system \cite{18Ye2018Fast,19Li2018Semi,20Wang2018Robust,21Sui2018Convex,22Zuo2017Distance,100chen2019data}. The key to metric learning is how to learn different metrics (such as Euclidean distance or other distance metrics) to represent the relationships between different entities instead of inner products. Metric learning is usually applied in the computer vision area, in which a deep transfer metric learning method for cross-domain visual recognition was proposed \cite{23Hu2015Deep}. For recommender systems, CML \cite{1hsieh2017collaborative} directly uses metric learning to embed the relationships between users and items, as shown in the upper part of Fig.\ref{Fig:MME2}. And IML \cite{6ijcai2018-389} proposes a practical framework to accelerate the embedding process. 
	
	Recently, some researchers combine metric learning with other existing models to improve performance. Combined with multi-task learning, CRML \cite{112DBLP:journals/nn/WuZNC20} is a metric learning model proposed for collaborative recommendations with co-occurrence embedding regularization. It considers the optimization problem as a multi-task learning problem which includes optimizing a primary task of metric learning and two auxiliary tasks of representation learning. To combine different styles of loss functions, SML \cite{121DBLP:conf/aaai/LiZZQZHH20} symmetrically introduces a positive item centric metric which maintains a closer distance from positive items to the user and pushes the negative items away from the positive items at the same time with an adaptive margin. Few researches focus on how to utilize metric learning to embed matrix, which is an open issue in the recommender system area.
	
	\subsection{Neural-network based recommendation models (NN-RSs)}
	The combination of recommender systems and the neural network is becoming a hot research trend \cite{19Li2018Semi,104DBLP:conf/www/GeWWQH20,12zhang2017deep,125DBLP:journals/tkde/HeHSLJC18}. Researchers attempt to utilize the non-linear activation functions in the neural network to measure the relationships between users and reviews. \cite{5he2017neural} utilizes a Multilayer perceptron (MLP) to design a network NeuCF to tackle implicit feedback recommendation problems. NeuCF is a rating-based model that can cover basic MF and CF and also achieve state-of-the-art performance. \cite{121DBLP:conf/aaai/LiZZQZHH20} combines semi-supervised and neural networks, bridges them, and reinforces mutually. 
	
	To tackle the sparse data in real-world scenarios, most existing neural-network based models use two-stage framework: first, it employs the matrix embedding or other models to embed the data into vectors. Then they input these vectors to achieve recommendations \cite{123DBLP:conf/www/Yuan0JGXXX20,124DBLP:conf/www/ChenZMLM20,5he2017neural,118DBLP:journals/tkde/ShiHZY19,117DBLP:conf/www/MaZYYCTHC20,120DBLP:conf/www/0003W0HC19}. \cite{117DBLP:conf/www/MaZYYCTHC20} proposes 2IPS, which is a two-stage off-policy policy gradient method. The proposed method explicitly takes into account the ranking model when training the candidate generation model, which helps improve the performance of the whole system. KTUP \cite{120DBLP:conf/www/0003W0HC19} jointly learns the model of recommendation and knowledge graph completion by combining several transfer schemes. It is an embedding-based recommender model with matrix embeddings. HERec \cite{118DBLP:journals/tkde/ShiHZY19} is a heterogeneous network embedding based approach for heterogeneous information network (HIN) based recommendation. It is a path-based recommender model with matrix embeddings. Also some GNN based models \cite{125DBLP:journals/tkde/HeHSLJC18,126DBLP:conf/sigir/Wang0WFC19,127DBLP:conf/www/Fan0LHZTY19,128,129,130,131DBLP:conf/sigir/0001DWLZ020} are boosting recently, including GraphRec \cite{127DBLP:conf/www/Fan0LHZTY19}, NGCN \cite{126DBLP:conf/sigir/Wang0WFC19} and LRGCCF \cite{129}, which greatly improve the recommender system. 
	
	\subsection{Relations among ME, ML and NN-based RSs}
	
	NN-based RSs is an important branch of recommender systems, which utilizes the strong computing ability of neural networks. According to the structure of NN, matrix embedding should be employed to project the abundant information in the user-item matrix into latent vectors. In this paper, we argue that existing ME models are not sufficient, and propose a representation learning model MML, which utilizes the idea of metric learning to enhance the ME performance, and benefit the NN-based RSs.
	
	\section{Conclusion}
	The quality of matrix embedding is an imperceptible but important factor in achieving a good recommendation. In this paper, we propose a matrix embedding model: Magnetic Metric Learning, which utilizes dual triplets to embed users and items with a metric-based loss function. With this model, we can achieve a unified embedding in a unified latent vector space. Through the experimental results on four datasets, our model is proved to be superior not only when compared with state-of-the-art models on all evaluation metrics, but also when trying to find a more stable latent space with the consideration of accuracy, efficiency, and overfitting. Our future work is to apply MML with some context and side information about users and items, to construct a more reasonable similar-pair set for latent relationships.
	
	\scriptsize
	\section*{Acknowledgment}
	This work is supported by the National Natural Science Foundations of China under Grant No. 61772230, No.61976102, No.U19A2065, and No. 61972450, Natural Science Foundation of China for Young Scholars No. 61702215 and No. 62002132, China Postdoctoral Science Foundation No. 2020M681040 and Changchun Science, and Technology Development Project No.18DY005, and National Defense Science and Technology Key Laboratory Fund Project No. 61421010418 and Science Foundation of Jilin Province No. 20190201022JC and China National Postdoctoral Program for Innovative Talents No. BX20180140.
	{\scriptsize
		\bibliographystyle{IEEEtran}
		\bibliography{ijcaixu}} 
	\vspace{-40pt}
	\begin{IEEEbiography}[{\includegraphics[width=1in,height=1.25in,clip,keepaspectratio]{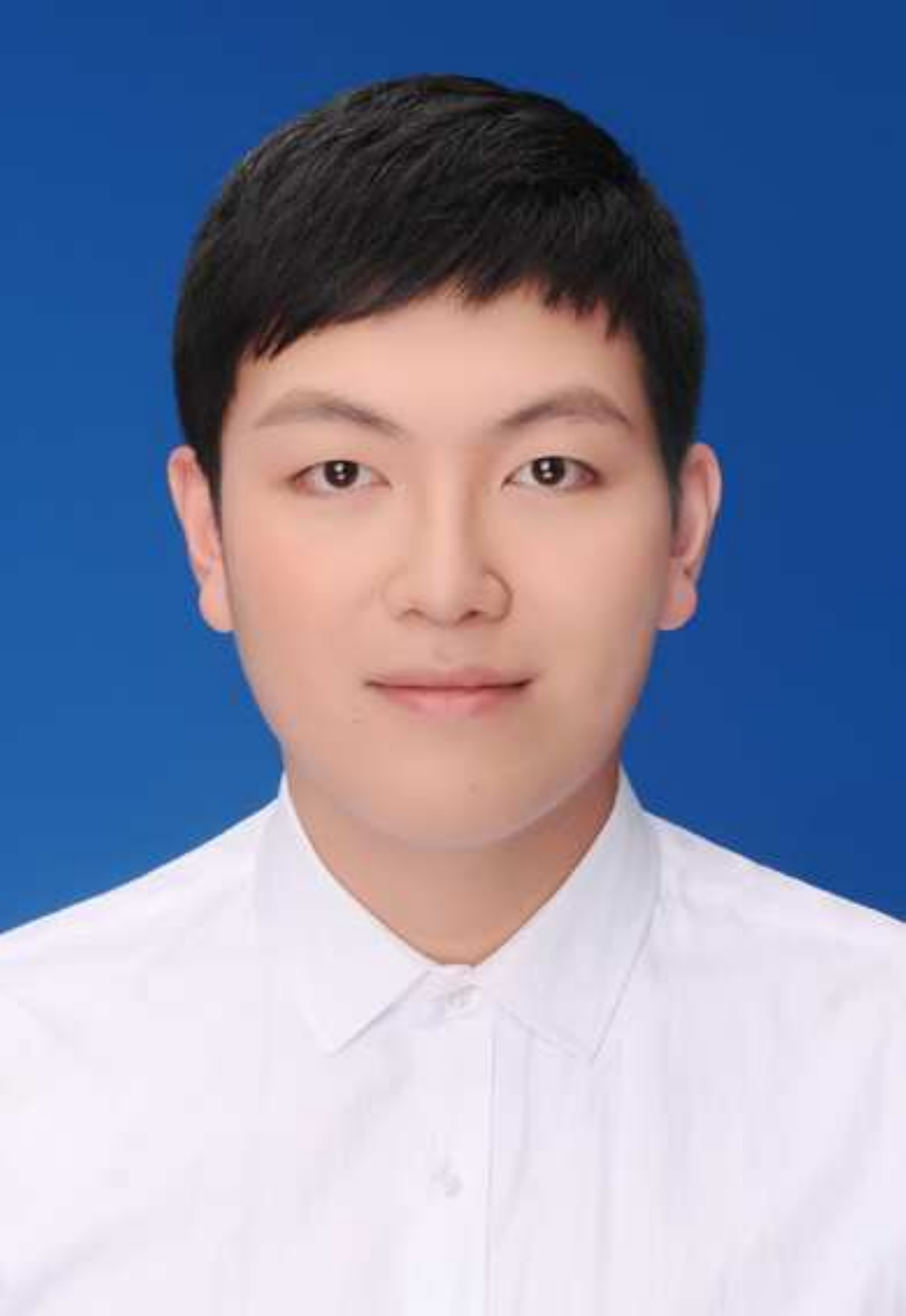}}]
		{Yuanbo Xu} received his B.E. degree in computer science and technology from Jilin University, Changchun, in 2012, his M.E. degree in computer science and technology from Jilin University, Changchun, in 2015, and his Ph.D. in computer science and technology from Jilin
		University, Changchun, in 2019. He is currently a Postdoc in the Department of Artificial Intelligence at Jilin University, Changchun. His research interests include applications of data mining, recommender system, and mobile computing. He has published some research results on journals such as TMM, TKDD, TNNLS and conference as ICDM, SECON.
	\end{IEEEbiography}
	\vspace{-40pt}
	\begin{IEEEbiography}[{\includegraphics[width=1in,height=1.25in,clip,keepaspectratio]{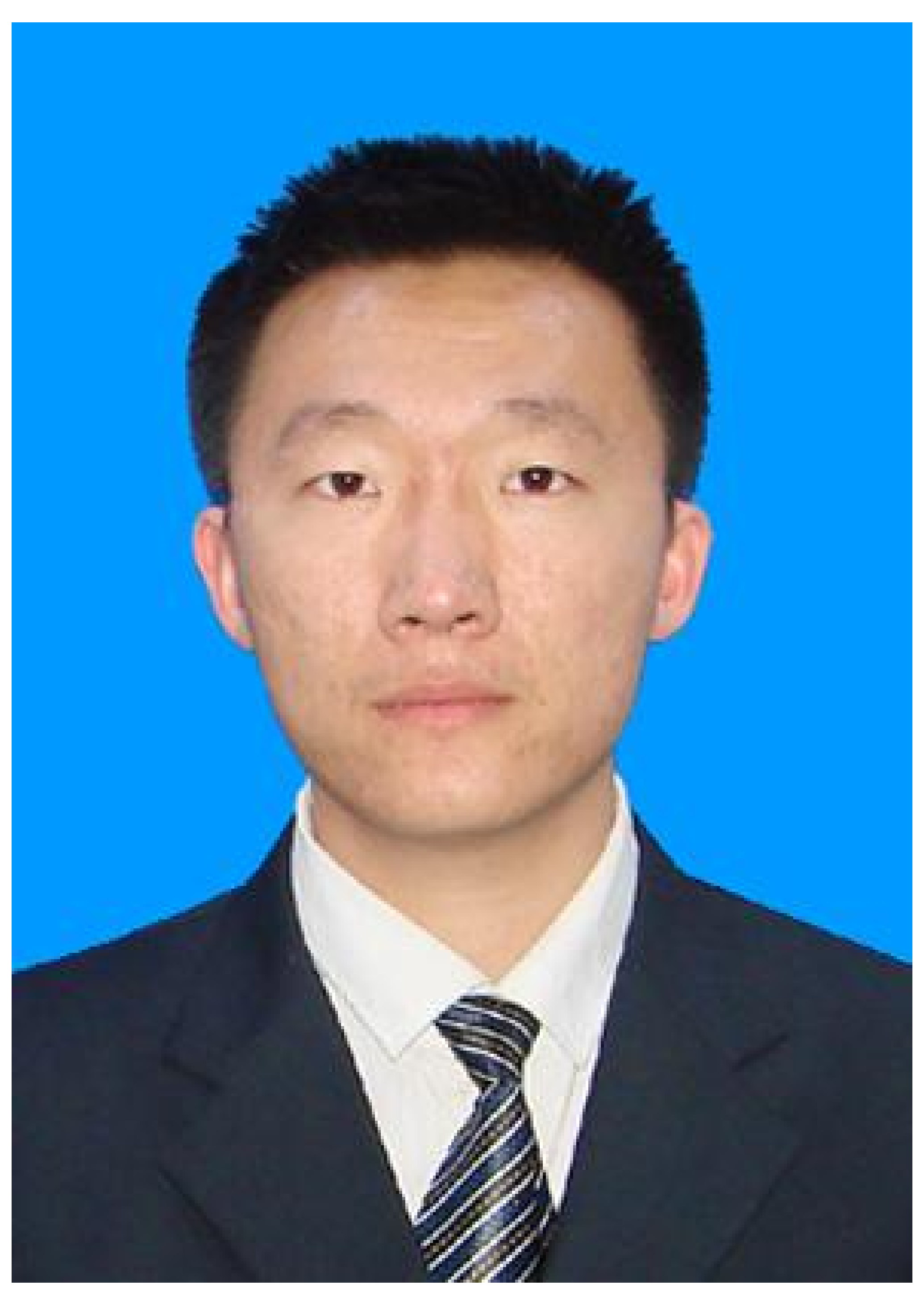}}]
		{En Wang} received his B.E. degree in software engineering from Jilin University, Changchun, in 2011, his M.E. degree in computer science and technology from Jilin University, Changchun, in 2013, and his Ph.D. in computer science and technology from Jilin
		University, Changchun, in 2016. He is currently an Associate Professor in the Department of Computer Science and Technology at Jilin University, Changchun. He is also a visiting scholar in the Department of Computer and Information Sciences at Temple University in Philadelphia. His current research focuses on the efficient utilization of network resources, scheduling and drop strategy in terms of buffer-management, energy-efficient communication between human-carried devices, and mobile crowdsensing.
	\end{IEEEbiography}
	\vspace{-40pt}
	\begin{IEEEbiography}[{\includegraphics[width=1in,height=1.25in,clip,keepaspectratio]{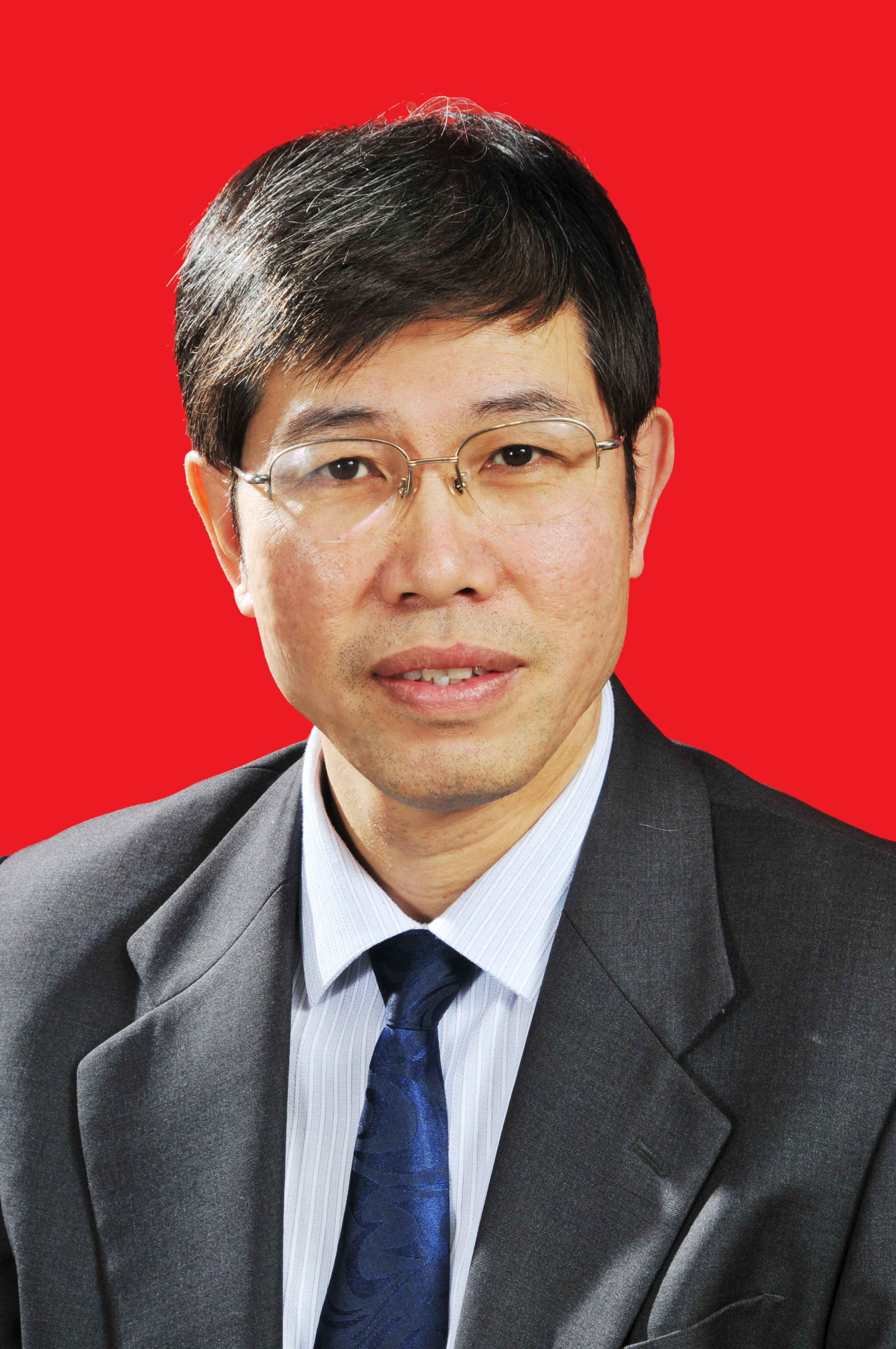}}]
		{Yongjian Yang} received his B.E. degree in automatization from Jilin University of Technology, Changchun, Jilin, China, in 1983; and M.E. degree in Computer Communication from Beijing University of Post and Telecommunications, Beijing, China, in 1991; and his Ph.D. in Software and theory of Computer from Jilin University, Changchun, Jilin, China, in 2005.  He is currently a professor and a PhD supervisor at Jilin University, Director of Key lab under the Ministry of Information Industry, Standing Director of Communication Academy, member of the Computer Science Academy of Jilin Province. His research interests include: Theory and software technology of network intelligence management; Key technology research of wireless mobile communication and services. He participated 3 projects of NSFC, 863 and funded by National Education Ministry for Doctoral Base Foundation. He has authored 12 projects of NSFC, key projects of Ministry of Information Industry, Middle and Young Science and Technology Developing Funds, Jilin provincial programs, ShenZhen, ZhuHai, and Changchun.
	\end{IEEEbiography}
	\vspace{-25pt}
	\begin{IEEEbiography}[{\includegraphics[width=1in,height=1.25in,clip,keepaspectratio]{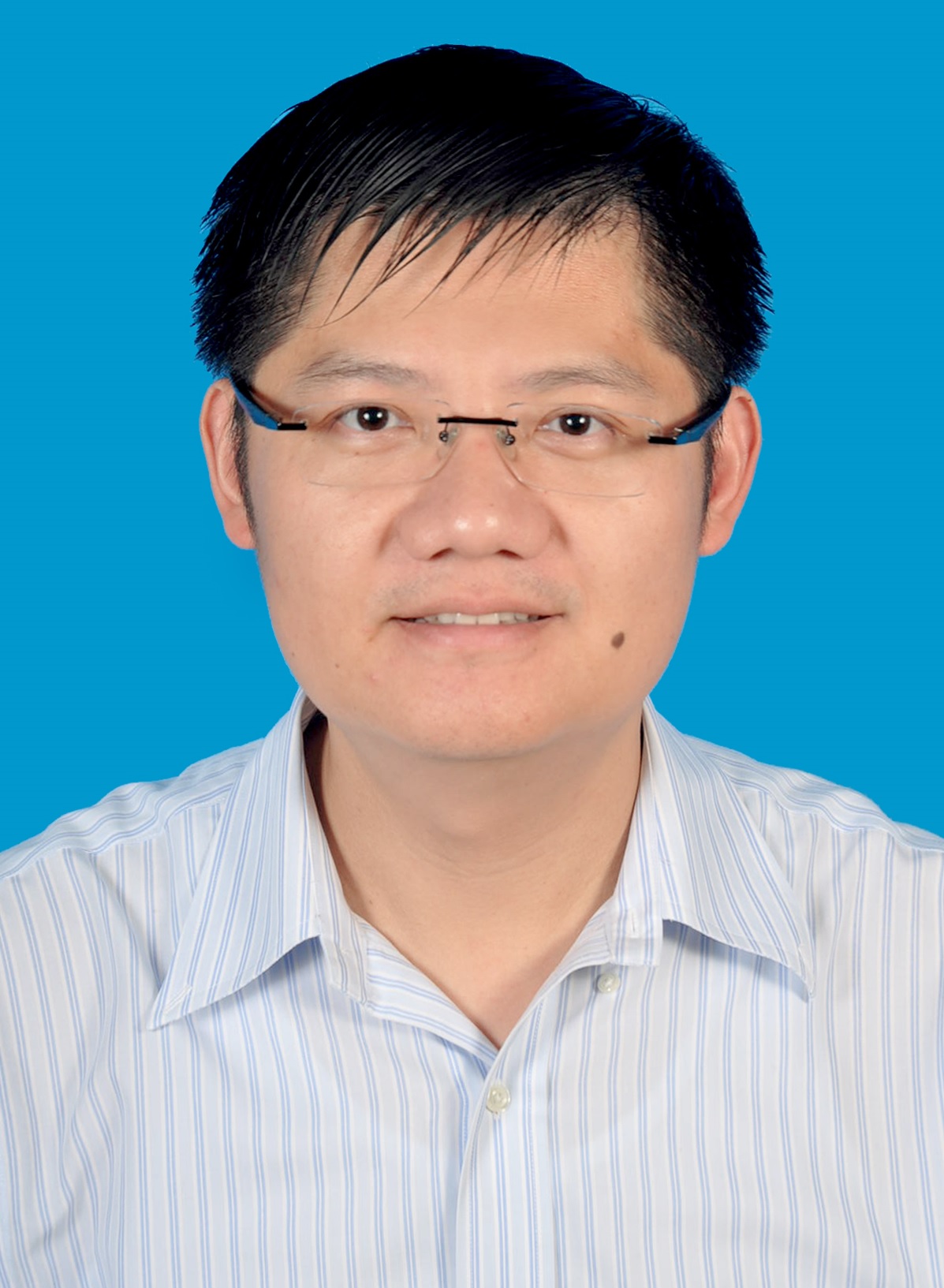}}]
		{Yi Chang} is dean of the School of Artificial Intelligence, Jilin University. His research interests include information retrieval, data mining, machine learning, natural language processing, and artificial intelligence. He is an associate editor of IEEE TKDE, and he served as one of the conference General Chairs for ACM WSDM'2018 and ACM SIGIR'2020. He is an IEEE Senior Member and ACM Distinguished Scientist.
	\end{IEEEbiography}
	
\end{document}